\documentclass[lettersize,journal]{IEEEtran}
\IEEEoverridecommandlockouts
\usepackage{cite}
\usepackage[tbtags]{amsmath}
\usepackage{amssymb,amsfonts}
\usepackage{algorithm}
\usepackage{algorithmic}
\usepackage{graphicx}
\usepackage{textcomp}
\usepackage{xcolor}
\usepackage{caption}
\usepackage{commath}
\usepackage{multirow}
\usepackage{subcaption}
\usepackage{url}

\DeclareMathOperator*{\argmax}{arg\,max}
\DeclareMathOperator*{\argmin}{arg\,min}
\DeclareMathOperator*{\dprime}{\prime \prime}

\def\BibTeX{{\rm B\kern-.05em{\sc i\kern-.025em b}\kern-.08em
    T\kern-.1667em\lower.7ex\hbox{E}\kern-.125emX}}
\begin{document}

\title{Learning-Based Downlink Power Allocation in Cell-Free Massive MIMO Systems}

\author{Mahmoud Zaher, Özlem Tuğfe Demir, \IEEEmembership{Member, IEEE,} Emil Björnson, \IEEEmembership{Fellow, IEEE,} and Marina Petrova, \IEEEmembership{Member, IEEE}%
\thanks{The authors are with the Division of Communication Systems, KTH Royal Institute of Technology, SE-164 40 Stockholm, Sweden. (e-mail: \{mahmoudz, ozlemtd, emilbjo, petrovam\}@kth.se).}

\thanks{M. Petrova is also with the Mobile Communications and Computing Group at RWTH Aachen University, 52062 Aachen, Germany.}
\thanks{This work was supported by the FFL18-0277 grant from the Swedish Foundation for Strategic Research.}}
%\thanks{\copyright~2022 IEEE. Personal use of this material is permitted. Permission from IEEE must be obtained for all other uses, in any current or future media, including reprinting/republishing this material for advertising or promotional purposes, creating new collective works, for resale or redistribution to servers or lists, or reuse of any copyrighted component of this work in other works.\\DOI: 10.1109/TWC.2022.3192203}

\maketitle

\begin{abstract}

This paper considers a cell-free massive multiple-input multiple-output (MIMO) system that consists of a large number of geographically distributed access points (APs) serving multiple users via coherent joint transmission. The downlink performance of the system is evaluated, with maximum ratio and regularized zero-forcing precoding, under two optimization objectives for power allocation: sum spectral efficiency (SE) maximization and proportional fairness. We present iterative centralized algorithms for solving these problems. Aiming at a less computationally complex and also distributed scalable solution, we train a deep neural network (DNN) to approximate the same network-wide power allocation. Instead of training our DNN to mimic the actual optimization procedure, we use a heuristic power allocation, based on large-scale fading (LSF) parameters, as the pre-processed input to the DNN. We train the DNN to refine the heuristic scheme, thereby providing higher SE, using only local information at each AP. Another distributed DNN that exploits side information assumed to be available at the central processing unit is designed for improved performance. Further, we develop a clustered DNN model where the LSF parameters of a small number of APs, forming a cluster within a relatively large network, are used to jointly approximate the power coefficients of the cluster.
\end{abstract}


\section{Introduction}

Since the wireless traffic is continuously growing, each new cellular network generation must find new ways to multiplex more devices and increase the spectral efficiency (SE) per device. The cornerstone technology of 5G is massive multiple-input multiple-output (MIMO) \cite{marzetta2016fundamentals,bjornson2017book,bjornson2017massive,nguyen2018optimal}, where each access point (AP) is equipped with a large number of active antennas to enable spatially multiplexing of a large number of user equipments (UEs) on the same time-frequency resource. Looking beyond 5G, cell-free massive MIMO is an emerging post-cellular infrastructure that is meant for rebuilding wireless networks to provide ubiquitous connectivity \cite{Ngo2017b,zhang2019cell}.
The key characteristic is that a large number of distributed APs jointly serve the UEs within a given coverage area without creating artificial cell boundaries \cite{buzzi2017downlink,bjornson2019making,chakraborty2020efficient,demir2021foundations}. This allows for an improvement in network connectivity and energy efficiency \cite{Ngo2018a}. Cell-free massive MIMO might become a foundation for beyond 5G technologies owing to its ability to utilize macro-diversity and multi-user interference suppression to provide almost uniform service to the UEs \cite{chakraborty2020efficient,interdonato2019scalability,zhang2020prospective}.

The main challenges of building a cell-free network infrastructure are the computational complexity of signal processing and the huge fronthaul requirements for information exchange between the APs \cite{interdonato2019ubiquitous,burr2018cooperative,buzzi2019user,masoumi2019performance}. Hence, it is desirable to build the network using limited-capacity fronthaul links between the APs and the central processing unit (CPU) \cite{burr2018cooperative,masoumi2019performance,bashar2020exploiting,femenias2020fronthaul}. In \cite{sarajlic2019fully,jeon2019decentralized}, decentralized schemes are investigated for MIMO systems, which can be extended to cell-free systems. Recent works \cite{interdonato2019scalability,bjornson2020scalable,demir2021foundations} have focused on the scalability aspects of various algorithms when the network size grows large. Since the different UEs are served simultaneously on the same time-frequency resource, power allocation plays an important role in limiting the multi-user interference and optimizing the network-wide performance \cite{bjornson2017book,zhao2020power}. It was noticed in \cite{Ngo2017b,nayebi2017precoding} that the network-wide downlink (DL)  max-min fairness problem is quasi-convex, and the authors solved it using general-purpose solvers.
More efficient dedicated algorithms for network-wide DL 
max-min fairness and sum-SE maximization were developed in \cite{chakraborty2020efficient}. However, the computational complexities of the algorithms grow polynomially with the number of APs and UEs, thus resulting in unscalable solutions that potentially violate real-time processing constraints.

An alternative solution to the power allocation problem is to employ the ``learn to optimize'' approach which makes use of the fact that deep neural networks (DNNs) can learn rich patterns and approximate complex function mappings \cite{goodfellow2016deep,lee2018deep,sun2018learning,sanguinetti2018deep,chakraborty2019centralized,zhang2019deep}. This allows for a significant reduction in computational complexity compared to solving the optimization problems with traditional methods, resulting in a solution that is real-time implementable. The price to pay is the extensive training required before the DNN becomes operational, however, this does not constitute a problem since it is performed offline \cite{zhao2020power}. This is in addition to the approximation error that is dependent on the structure of the data, the training procedure, and the selected network architecture that will be discussed later on. In the context of cellular massive MIMO, the authors in \cite{sanguinetti2018deep} train a feedforward DNN that uses the exact UE locations, which might not be easily obtainable in practice, as input to learning the optimal power coefficients maximizing the max-min fairness and proportional fairness (PF) in the DL, whereas \cite{van2020power} develops a DNN for joint pilot and data power control for sum-SE maximization in the uplink (UL). In \cite{zhao2020power}, a centralized deep convolutional neural network is designed to approximate the power coefficients for the max-min fairness objective in the cell-free massive MIMO DL. Moreover, \cite{bashar2020exploiting} develops another centralized deep learning solution for the sum-SE maximizing power control in the UL of a cell-free system, taking into account the limited-capacity fronthaul links between the APs and the CPU. In \cite{chakraborty2019centralized}, centralized and distributed DNNs were developed for the max-min fairness power allocation in a small cell-free massive MIMO setup. The large-scale fading (LSF) coefficients were used as input to both the centralized and distributed DNN implementations. Especially in the distributed case, the results show that the approximation performance was not satisfactory, likely due to two main factors: 1) the inputs were not pre-processed based on prior domain knowledge of what information is essential for power allocation; and 2) the total transmit power from each AP was not approximated by the DNN, which is essential in the distributed case.

In this paper, a distributed solution to the DL power allocation problem in a cell-free network is proposed. We first formulate the network-wide sum-SE and PF-based power allocation optimization problems.\footnote{Although the quasi-convex max-min fairness problem can be solved to global optimality, the solution is undesirable since all UEs will get the same SE as the weakest UE in the entire network.
The simulation results in \cite{Ngo2017b} were generated by terminating the bisection algorithm before convergence; that is, before the SEs of the strongest UEs have been brought down to the SE of the weakest UE. Hence, only an approximate version of the problem is solved.}
To solve these non-convex  problems, we employ a weighted minimum mean square error (WMMSE) algorithm inspired by the previous work \cite{chakraborty2020efficient}. It is worth mentioning that PF-based power allocation was not considered in \cite{chakraborty2020efficient}. We take an alternating direction method of multipliers (ADMM) approach to efficiently obtain the solution of the quadratically-constrained quadratic sub-problems of the WMMSE algorithms, as was done in \cite{chakraborty2020efficient}. This allows for generating a large amount of training data in a reasonable  time. We then train a feedforward DNN to approximate the respective power coefficients. The recent line of work \cite{sun2018learning,lee2018deep,nasir2019multi,zhang2020multi,de2018team} utilizing distributed learning-based solutions for power allocation with either $n$ parallel interfering links (i.e., transceiver pairs in what information theorists call an interference channel) or interference management in (heterogeneous) cellular setups where each UE is only served by one transmitter, tackle fundamentally different communication scenarios. Machine learning can be applied in such setups to learn the inter-cell interference levels so that each AP can make autonomous power allocation decisions within its cell. This stands in sharp contrast to a cell-free massive MIMO setup, where each UE is served by a multitude of APs. In our proposed approach, we will not only learn the interference statistics, but also how the APs ``share the burden'' of serving the UEs. One clear indicator of this is that the number of power coefficients to be approximated in the above-mentioned papers is much smaller and the inter-relations between the coefficients are expected to be much simpler, since each receiver (UE) is only served by a single transmitter in contrast to a cell-free network where, for example, cutting down on one AP's power to reduce interference on a certain UE can be compensated by increasing the power from another AP. Moreover, the previously considered cellular scenarios are likely to have a closed-form WMMSE or fractional programming solution to the optimization objectives. However, in our case the computational complexity for solving the optimization problems using the WMMSE algorithm centrally is impractical for a real-time implementation. This is the first paper to provide distributed DNN solutions to the sum-SE and PF power allocation problems in the context of cell-free massive MIMO, to the best of our knowledge.

The main contributions of the paper can be summarized as follows:
\begin{itemize}
\item We develop a fully distributed feedforward DNN for each AP to approximate the per-AP normalized power coefficients and the total transmitted power from the AP, using only local information available at the AP as input, however, utilizing the network-wide sum-SE and PF solutions in the training phase as the labeled output. Accordingly, each AP employs a DNN that implicitly learns the essential network structure and propagation environment. A heuristic closed-form power allocation, based on the LSF coefficients between the AP and the UEs, is used as the input to the DNN. The heuristic input is chosen to make the input/output relation of the DNN model less complex, by providing better scaling of the input data.
\item We develop another distributed DNN that utilizes side information in the form of the ratio of the LSF coefficients from a given AP, to the LSF coefficients from all APs in the network; to enhance the performance of the learning-based power allocation. We further implement a clustered DNN where the LSF parameters of a small number of APs, forming a cluster within a relatively large network, is fed to the DNN to simultaneously learn the ``optimal" power allocation of the cluster. Compared to a distributed solution, this architecture provides better performance at the expense of increased fronthaul requirements.
\item We perform a complexity analysis of the required processing and run-time for the proposed DNN models in comparison to the conventional methods of solving the power allocation optimization problems.
\end{itemize}

It is worth noting that the power allocation problem is more complicated in the DL than in the UL. This is due to 1) more power coefficients need to be approximated in the DL; and 2) the power constraints are more involved in the DL since the power budget of each AP is distributed among the UEs, whereas in the UL each UE has his own separate power budget.

The rest of the paper is organized as follows: Section \ref{system} outlines the system model. The sum-SE and PF maximization problems are presented in Section \ref{optimization}. Sections \ref{secDist} and \ref{secClu} discuss the proposed distributed and clustered DNN-based power allocation, respectively. Further details about the DNN structures and complexity analysis are provided in Section \ref{complexity}. Section \ref{numerical} presents numerical results, whereas the main conclusions of the paper are stated in Section \ref{conc}.

\subsection{Notations}

Lowercase and uppercase boldface letters denote column vectors and matrices, respectively. $\mathcal{N}_{\mathbb{C}}(\mathbf{0}, \mathbf{R})$ denotes the distribution of a multivariate circularly symmetric complex Gaussian variable with zero mean and covariance matrix $\mathbf{R}$. The symbols $(\cdot)^T$, $(\cdot)^H$, and $(\cdot)^{-1}$ are used to indicate transpose, conjugate transpose, and the inverse of a matrix, respectively. The expectation, trace, and L$_2$ vector norm are denoted by $\mathbb{E}(\cdot)$, tr$(\cdot)$ and $\norm{\cdot}$, respectively. $\mathbf{I}_M$ represents the $M \times M$ identity matrix.

\subsection{Reproducible research}

All the simulation results can be reproduced using the Python code and data files available at: https://github.com/emilbjornson/power-allocation-cell-free.

\section{System Model}\label{system}

This paper considers a cell-free massive MIMO system with $K$ single-antenna UEs that are arbitrarily distributed in a large service area.
The UEs are jointly served by $L$ APs, each equipped with $N$ antennas. We assume a standard block fading channel model where the time-varying wideband channels are divided into time-frequency coherence blocks such that the channels are static and frequency-flat in each block \cite{bjornson2017book}. The coherence block is comprised of $\tau_c$ symbols and the channels take independent random realizations in each coherence block. The channel between UE $k$ and AP $l$ is represented by $\mathbf{h}_{kl} \in \mathbb{C}^{N \times 1}$ and is modelled by correlated Rayleigh fading as $\mathbf{h}_{kl} \sim \mathcal{N}_{\mathbb{C}}(\mathbf{0}, \mathbf{R}_{kl})$, where $\mathbf{R}_{kl} \in \mathbb{C}^{N \times N}$ denotes the spatial correlation matrix. The average channel gain from a given antenna at AP $l$ to UE $k$ is determined by the normalized trace $\beta_{kl} = \frac{1}{N} \hspace{1pt}\textrm{tr}\hspace{-1pt}\left(\mathbf{R}_{kl}\right)$.

\begin{figure}
\centering
\setlength{\abovecaptionskip}{0.3cm}
\includegraphics[scale=0.66]{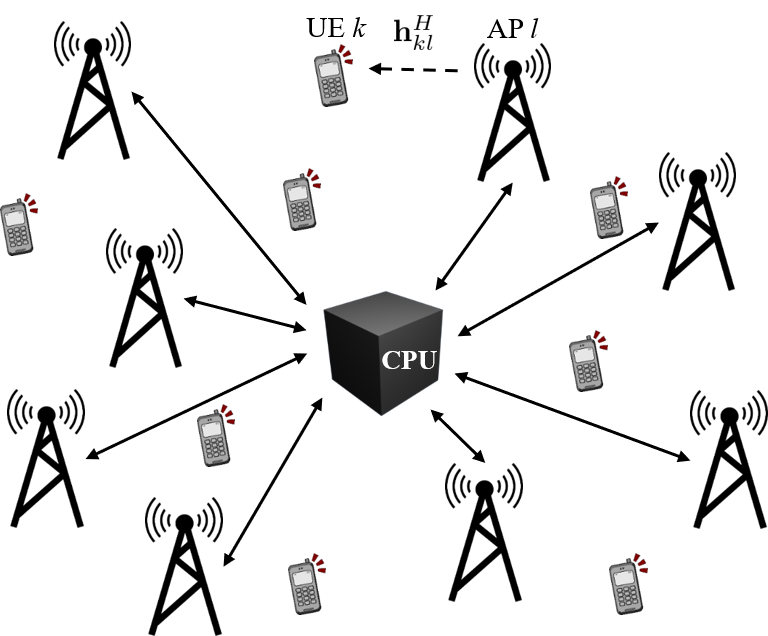}
\caption{Cell-free network architecture. The solid lines represent fronthaul links connecting the APs to the CPU.}
\label{model1}
\end{figure}

Fronthaul links connect the APs to a CPU, conveying UL and DL data between them and other necessary signals. The connections are assumed to be error-free, and no instantaneous channel state information (CSI) is transmitted over the fronthaul, but only data and channel statistics  \cite{bjornson2019making,nayebi2017precoding}. Fig.~\ref{model1} depicts the network architecture. There are two main ways of operating such a network: centralized and distributed operation \cite{demir2021foundations}.
We consider the distributed operation, where each AP selects the precoding vectors based on the locally estimated channels. In the network-wide power allocation schemes that are conventionally used in cell-free massive MIMO systems, the CPU is responsible to select the power allocated to each UE by each AP using the channel statistics that are shared from the APs. The rate with which new power allocation decisions need to be made and communicated to the APs is determined by the users' large-scale mobility and scheduling decisions in the network. Optimally, the system must be capable of making new power allocation decisions at the same frequency as the scheduling decisions can change, which happens at the subframe level (1\,ms) in 5G. Large-scale UE mobility also requires changes in power allocation but this happens more slowly. Ideally, the transmit powers should also be selected locally at each AP and this paper develops efficient algorithms for that purpose.

\subsection{Channel Estimation}

We consider a time-division duplex (TDD) protocol that consists of a pilot transmission phase for channel estimation and a data transmission phase. Following \cite{bjornson2017book}, the coherence block is divided into three parts: $\tau_p$ samples for UL pilots, $\tau_u$ samples for UL data transmission, and $\tau_d$ samples for DL data transmission. Accordingly, we have $\tau_c = \tau_p + \tau_u + \tau_d$. Since this paper focuses on DL power allocation, only UL pilot and DL data transmission are considered, i.e., $\tau_u=0$.

In the channel estimation phase, each UE is assigned a random $\tau_p$-length pilot from a set of $\tau_p$ mutually orthogonal pilots utilized by the APs. Let  $t_k\in \{1,\hdots,\tau_p\}$ denote the index of the pilot assigned to UE $k$. After correlating the received signal at AP $l$ with pilot $t_k$ for the estimation of the UE $k$ channel, the signal $\mathbf{y}_{t_kl}^p \in \mathbb{C}^{N \times 1}$ is obtained as
\begin{equation}
\mathbf{y}_{t_kl}^p = \sum_{\substack{i = 1\\ t_i = t_k}}^{K}\sqrt{\tau_pp_i}\mathbf{h}_{il} + \mathbf{n}_{t_kl},
\end{equation}
where $p_i$ is the transmit power of UE $i$ and $\mathbf{n}_{t_kl} \hspace{-0.85pt}\sim \mathcal{N}_{\mathbb{C}}\hspace{-0.5pt}\left(\mathbf{0}, \sigma^2\mathbf{I}_N\hspace{-0.5pt}\right)$ represents the additive Gaussian noise vector at AP $l$. Employing the minimum mean square error (MMSE) estimator at AP $l$, the channel between UE $k$ and AP $l$ is estimated as \cite{demir2021foundations}
\begin{equation}
\begin{split}
\hat{\mathbf{h}}_{kl} &= \sqrt{\tau_pp_k}\mathbf{R}_{kl}\left(\sum_{\substack{i = 1 \\ t_i = t_k}}^{K}\tau_pp_i\mathbf{R}_{il} + \sigma^2\mathbf{I}_N\right)^{\hspace{-1pt}-1}\hspace{-1pt}\mathbf{y}_{t_kl}^p\\
&\sim \mathcal{N}_{\mathbb{C}}\left(\mathbf{0}, \tau_pp_k\mathbf{R}_{kl}\mathbf{\Psi}_{kl}^{-1}\mathbf{R}_{kl}\right)
\end{split}
\end{equation}
where $\mathbf{\Psi}_{kl} = \mathbb{E}\left\{\mathbf{y}_{t_kl}^p\left(\mathbf{y}_{t_kl}^p\right)^H\right\} = \sum_{i=1, t_i = t_k}^K\tau_pp_i\mathbf{R}_{il} + \sigma^2\mathbf{I}_N$ represents the correlation matrix of the received pilot signal.

\subsection{Pilot Assignment}\label{pAssign}

For a UE to get access to the network, it needs to be assigned to a pilot sequence. To focus on the power allocation problem, the low-complexity assignment algorithm from \cite{demir2021foundations} is adopted, where the first $\tau_p$ UEs are assigned orthogonal pilots. The remaining UEs are assigned to the pilots as follows; for UE $k$, $k = \tau_p + 1, \hdots, K$:

{\bf Step 1:} The UE determines the AP to which it has the strongest signal and appoints it as its \textit{Master AP}. The index of this AP is given by
\begin{equation}
l^{\star} = \argmax_{l \in \{1, \hdots, L\}} \beta_{kl}.
\end{equation}

{\bf Step 2:} The \textit{Master AP} identifies the pilot index where the respective pilot interference from the first $k-1$ UEs (that has already been assigned a pilot) is minimal, i.e.,
\begin{equation}
\tau = \argmin_{t \in \{1, \hdots, \tau_p\}} \sum_{\substack{i = 1 \\ t_i = t}}^{k-1}\beta_{il^{\star}}.
\end{equation}

This pilot is assigned to UE $k$, and the algorithm then continues with the next UE. In a nutshell, each UE is sequentially assigned to the pilot sequence where it experiences the least pilot contamination. The algorithm will effectively avoid the worst-case situations where two co-located UEs are assigned to the same pilot. This reduces the ambiguity in the training dataset for the NNs in situations where the UEs have similar LSF configurations, but different channel estimation accuracy due to pilot assignments.

\subsection{Downlink Data Transmission}

All UEs are assumed to be served by all the $L$ APs on the same time-frequency resources. As a result, the received DL signal at UE $k$ is given by
\begin{equation}
y_k^{dl} = \sum_{l = 1}^{L}\mathbf{h}_{kl}^H\sum_{i = 1}^{K}\sqrt{\rho_{il}}\mathbf{w}_{il}s_i + n_k,
\end{equation}
where $\rho_{il}\geq 0$ is the allocated DL power by AP $l$ to UE $i$ and $\mathbf{w}_{il} \in \mathbb{C}^{N \times 1}$ is the corresponding normalized precoding vector, such that $\norm{\mathbf{w}_{il}}^2 = 1$. Moreover, $s_i$ denotes the zero-mean signal intended for UE $i$ and $n_k \sim \mathcal{N}_{\mathbb{C}}\hspace{-1pt}\left(0, \sigma^2\right)$ represents the noise at UE $k$.

A lower bound on the ergodic DL SE in a cell-free system is given by the following lemma.

\noindent\textbf{Lemma 1}\cite[Lem.~1]{chakraborty2020efficient} \textit{The achievable SE for UE $k$ is lower bounded by}
\begin{equation}
\textrm{SE}_k = \frac{\tau_d}{\tau_c}\textrm{log}_2\left(1 + \textrm{SINR}_k\right) \label{eq:SEk}
\end{equation}
\textit{where}
\begin{equation}
\textrm{SINR}_k = \frac{\left(\mathbf{a}_k^T\boldsymbol{\mu}_k\right)^2}{\sum_{i = 1}^{K}\boldsymbol{\mu}_i^T\mathbf{B}_{ki}\boldsymbol{\mu}_i - \left(\mathbf{a}_k^T\boldsymbol{\mu}_k\right)^2 + \sigma^2}
\label{SINR}
\end{equation}
\textit{represents the effective SINR and}
\begin{align}
&\boldsymbol{\mu}_k = \left[\mu_{k1} \cdots \mu_{kL}\right]^T \in \mathbb{R}^{L \times 1},\ \mu_{kl} = \sqrt{\rho_{kl}} \label{eq:mu}\\
&\mathbf{a}_k = \left[a_{k1} \cdots a_{kL}\right]^T \in \mathbb{R}^{L \times 1},\ a_{kl} = \mathbb{E}\left\{\mathbf{h}_{kl}^H\mathbf{w}_{kl}\right\}\\
&\mathbf{B}_{ki} \in \mathbb{R}^{L \times L},\ b_{ki}^{lm} = \Re\left(\mathbb{E}\left\{\mathbf{h}_{kl}^H\mathbf{w}_{il}\mathbf{w}_{im}^H\mathbf{h}_{km}\right\}\right)
\end{align}
\textit{and} $b_{ki}^{lm}$ \textit{denotes element (l,m) in matrix} $\mathbf{B}_{ki}$.
\vspace{2pt}

In \eqref{SINR}, $\mathbf{a}_k^T\boldsymbol{\mu}_k$ corresponds to the desired signal gain for UE $k$ over deterministic precoded channel. The SE expression is applicable for any precoding scheme, with precoding vectors rotated such that $a_{kl} = \mathbb{E}\left\{\mathbf{h}_{kl}^H\mathbf{w}_{kl}\right\}$ is real and non-negative, and the considered correlated Rayleigh fading model. The considered precoding vectors $\{\mathbf{w}_{il}\}$ satisfy short-term power constraints, meaning that $\norm{\mathbf{w}_{il}}^2 = 1$ is satisfied in each coherence block, and not on average. The relaxation of this condition such that $\mathbb{E}\{\norm{\mathbf{w}_{il}}^2\} = 1$ is not preferable in cell-free massive MIMO as the channel between an AP with few antennas and a UE does not harden \cite{chakraborty2019centralized}. The normalized precoding vector is defined as $\mathbf{w}_{il} = \bar{\mathbf{w}}_{il}/\norm{\bar{\mathbf{w}}_{il}}$, such that $\bar{\mathbf{w}}_{il}$ can be chosen arbitrarily. 

The main contributions of this work are applicable along with any precoding scheme. However, in the numerical evaluation, we will employ the maximum ratio (MR) and regularized zero-forcing (RZF) precoding schemes defined as
\begin{equation}
\bar{\mathbf{w}}_{kl} =
\begin{cases}
      \hat{\mathbf{h}}_{kl} & \textrm{for MR}, \\
      \left(\sum\limits_{i = 1}^{K}p_i\hat{\mathbf{h}}_{il}\hat{\mathbf{h}}_{il}^H + \sigma^2\mathbf{I}_N\right)^{-1}p_k\hat{\mathbf{h}}_{kl} & \textrm{for RZF}.
\end{cases}
\end{equation}

\section{Sum-SE and PF Maximizing Power Allocation} \label{optimization}

In this section, we formulate the sum-SE and PF maximizing power allocation problems. The aim is to find the DL power allocation coefficients $\{\rho_{kl}: \forall k, l\}$, that maximize the sum-SE or product-SE (in the case of PF) under the same per-AP power constraints
\begin{equation}
\sum_{k = 1}^{K}\rho_{kl} \leq P_{\textrm{max}}^{\textrm{dl}}, \quad l = 1, \hdots, L, 
\end{equation}
where $P_{\textrm{max}}^{\textrm{dl}}$ is the maximum allowable transmit power for each AP. To expose a hidden problem structure, we express both problems in terms of the variables $\{\mu_{kl}\}$ in \eqref{eq:mu}, which are the square roots of the power allocation coefficients $\{\rho_{kl}\}$. Since these variables only appear in quadratic forms, we do not include the constraints $\mu_{kl}\geq 0$ in the considered optimization problems (similar to \cite{chakraborty2020efficient}) as this allows us to obtain closed-form update equations in the resulting algorithms. Note that this relaxation does not lead to any issue since the sign of any negative $\mu_{kl}$ at the end of the optimization algorithms can be replaced by the positive counterpart  without invalidating any other constraints.

\subsection{Sum-SE Maximization}

Removing the constant pre-log factor in \eqref{eq:SEk}, the sum-SE maximizing power allocation problem can be expressed as 
\begin{equation}
\begin{split}
\mathop{\mathrm{maximize}}\limits_{\{\mu_{kl}: \forall k, l\}} \quad &\sum_{k = 1}^{K}\textrm{log}_2\left(1 + \textrm{SINR}_k\right)\\
\textrm{subject to} \quad &\sum_{k = 1}^{K}\mu_{kl}^2 \leq P_{\textrm{max}}^{\textrm{dl}}, \quad l = 1, \hdots, L.
\label{sumse}
\end{split}
\end{equation}
\begin{algorithm}[t]
	\caption{WMMSE algorithm for solving sum-SE and PF problems}
	\label{alg1}
	\noindent\textbf{Input:} Initialize optimization variables $\mu_{kl}^{(0)}$, $\forall k,l$. Set the iteration index $n\leftarrow 0$ and the solution accuracy \\$\epsilon_{\rm WMMSE}>0$.
	\begin{algorithmic}[1]
		\REPEAT
		\STATE Update the variables $v_{k}^{(n+1)}$, $k=1,\ldots,K$  as
		\begin{equation}\label{eq:Updated vk}
		v_{k}^{(n+1)} = \frac{ \mathbf{a}_{k}^{T}\boldsymbol{\mu}_{k}^{(n)}}{\sum_{i=1}^{K}\left(\boldsymbol{\mu}_{i}^{(n)}\right)^{T} \mathbf{B}_{ki}\boldsymbol{\mu}_{i}^{(n)} +  \sigma^{2}}.
		\end{equation}
		\STATE Update the variables $e_k^{(n+1)}$, $k=1,\ldots,K$ as
		\begin{equation}\label{eq:Updated error}
		e_{k}^{(n+1)}= 1 - \frac{\left(\mathbf{a}_{k}^{T} \boldsymbol{\mu}_{k}^{(n)}\right)^{2}}{\sum_{i=1}^{K}\left(\boldsymbol{\mu}_{i}^{(n)}\right)^{T} \mathbf{B}_{ki}\boldsymbol{\mu}_{i}^{(n)} +  \sigma^{2}}.
		\end{equation}
		\STATE Update the variables $\omega_{k}^{(n+1)}$, $k=1,\ldots,K$ as \\
		\emph{if sum-SE maximization:}
		\begin{equation}\label{eq:Updated weight0}
		\omega_{k}^{(n+1)} = 1\Big/e_{k}^{(n+1)}, 
		\end{equation}
		\emph{if PF maximization:}
		\begin{equation}\label{eq:Updated weight}
		\omega_{k}^{(n+1)} =
	-1\Big/\left(e_{k}^{(n+1)}\ln\left(e_{k}^{(n+1)}\right)\right).
		\end{equation}
		\STATE Update $\left\{\mu_{kl}^{(n+1)}: \forall k,l\right\}$ as the solution to the problem
		\begin{align}
&\mathop{\mathrm{minimize}}\limits_{\{\mu_{kl}: \forall k, l\} }\,\,  \quad \sum_{k=1}^{K}  \omega_{k}^{(n+1)}\Bigg[\left(v_{k}^{(n+1)}\right)^{2} \nonumber\\
& \times \Bigg(\sum_{i=1}^{K}\boldsymbol{\mu}_{i}^{T} \mathbf{B}_{ki}\boldsymbol{\mu}_{i}+  \sigma^{2}\Bigg) - 2 v_{k}^{(n+1)}\mathbf{a}_{k}^{T} \boldsymbol{\mu}_{k} + 1 \Bigg] \nonumber \\
&\textrm{subject to}  \quad  \sum\limits_{k=1}^{K}\mu_{kl}^{2}\leq P^{\mathrm{dl}}_{\textrm{max}}, \quad  l=1,\ldots,L.\label{eq:power constraint}
\end{align}
		\STATE Set $n\leftarrow n+1$.
		\UNTIL{The square of the difference between the objective functions at the current and the previous iteration is less than  $\epsilon_{\rm WMMSE}$.}
	\end{algorithmic}
	\textbf{Output:} The local optima: $\mu_{kl}^{\star} = \mu_{kl}^{(n)}: \forall k,l$.
\end{algorithm}

The problem \eqref{sumse} is non-convex and finding the global optimal solution is highly complicated. In this paper, we will use the WMMSE algorithm to obtain a local optimum solution with an efficient iterative algorithm, which is important if the algorithm should be used to generate a large training dataset. A similar Monte Carlo methodology to \cite{sanguinetti2018deep} has been adopted to compute the optimal powers. The main steps of the algorithm are outlined in Algorithm~\ref{alg1} for the sake of completeness. Note that all the steps of the algorithm, except for Step 4, are the same for both sum-SE and PF problems; so we present both of them in Algorithm~\ref{alg1}. The superscript $^{(n)}$ denotes the values at iteration $n$. The auxiliary variables $v_{k}^{(n)}$, $e_k^{(n)}$, and $\omega_k^{(n)}$ in Algorithm~\ref{alg1} come from the WMMSE reformulation \cite{chakraborty2020efficient,christensen2008weighted, shi2011iteratively}. The sub-optimization problem in Step 5 of the algorithms is a convex quadratically-constrained quadratic programming problem and it can be solved with the closed-form ADMM algorithm in \cite[Alg.~1]{chakraborty2020efficient}.

\subsection{PF Maximization}

The sum-SE maximization problem considered in the previous part does not consider any fairness among the UEs. In fact, more emphasis is given to the relatively fortunate UEs in the network in an effort to maximize the sum-SE. An alternative to the sum-SE utility is PF \cite{shi2011iteratively, diamantoulakis2017maximizing,chen2019proportional}, which can be formulated as the sum of logarithms of the individual SE of the UEs. PF maximization 
obtains a good balance between max-min fairness, which limits the network-wide performance by focusing on the most unfortunate UEs with the worst channel conditions, and sum-SE maximization. The respective optimization problem is expressed as
\begin{equation}
\begin{split}
\mathop{\mathrm{maximize}}\limits_{\{\mu_{kl}: \forall k, l\}} \quad &\sum_{k = 1}^{K}\textrm{ln}\left(\textrm{log}_2\left(1 + \textrm{SINR}_k\right)\right)\\
\textrm{subject to} \quad &\sum_{k = 1}^{K}\mu_{kl}^2 \leq P_{\textrm{max}}^{\textrm{dl}}, \quad l = 1, \hdots, L
\label{pf}
\end{split}
\end{equation}
where the constant pre-log factors have been omitted as they do not affect the optimization problem. Similar to the sum-SE maximization problem, the PF problem is also non-convex and it is hard to obtain the global optimum solution with a reasonable complexity. The problem structure for PF maximization poses a different reformulation to obtain closed-form updates for the auxiliary variables introduced by the WMMSE algorithm. First, an estimate of the data symbol of UE $k$ is obtained by applying the receiver weight $v_k$ as
\begin{equation}
\hat{s}_k = v_ky^{dl}_k = v_k\sum_{l = 1}^{L}\mathbf{h}_{kl}^H\sum_{i = 1}^{K}\sqrt{\rho_{il}}\mathbf{w}_{il}s_i + v_kn_k.
\end{equation}
Then, defining the mean square error (MSE) as
\begin{equation}
\begin{split}
e_k &= \mathbb{E}\left\{|\hat{s}_k - s_k|^2\right\}\\
&= v_k^2\left(\sum_{i=1}^K\boldsymbol{\mu}_i^T\mathbf{B}_{ki}\boldsymbol{\mu}_i + \sigma^2\right) - 2v_k\mathbf{a}_k^T\boldsymbol{\mu}_k + 1,\\
\end{split}
\end{equation}
we can reformulate the problem in \eqref{pf} as
\begin{equation}
\begin{split}
\mathop{\mathrm{minimize}}\limits_{\substack{\{\mu_{kl}: \forall k, l\}\\ \{\omega_k \geq 0, v_k: \forall k\}}} \quad &\sum_{k = 1}^{K}\left(\omega_ke_k + f\hspace{-1pt}\left(g\hspace{-1pt}\left(\omega_k\right)\right) - \omega_kg\hspace{-1pt}\left(\omega_k\right)\right)\\
\textrm{subject to} \quad &\sum_{k = 1}^{K}\mu_{kl}^2 \leq P_{\textrm{max}}^{\textrm{dl}}, \quad l = 1, \hdots, L.
\label{pf2}
\end{split}
\end{equation}

The function $f(x)$ is defined as $f(x) = -\textrm{ln}(-\textrm{ln}(x))$ for $0 < x < 1$. Its derivative can be written as $f^\prime(x) = -1/(x\textrm{ln}(x))$, which is positive for the defined range of $x$. The function $g(x)$ is the right inverse function of $f^\prime(x)$, i.e., $f^\prime(g(x)) = x$.

The equivalence of the problems \eqref{pf} and \eqref{pf2} in terms of having the same global optimum follows from the fact that the optimum $\omega_k$ is obtained by equating the derivative of the objective in \eqref{pf2} to zero, resulting in
\begin{equation}
e_k + f^\prime\hspace{-1pt}\left(g\hspace{-1pt}\left(\omega_k\right)\right)g^\prime\hspace{-1pt}\left(\omega_k\right) - g\hspace{-1pt}\left(\omega_k\right) - \omega_kg^\prime\hspace{-1pt}\left(\omega_k\right) = 0.
\end{equation}
Since $f^\prime\hspace{-1pt}\left(g\hspace{-1pt}\left(\omega_k\right)\right) = \omega_k$, we have $e_k = g\hspace{-1pt}\left(\omega_k\right)$, and accordingly $\omega_k^{\textrm{opt}} = f^\prime\hspace{-1pt}\left(e_k\right)$. Plugging $\omega_k^{\textrm{opt}}$ into the objective in \eqref{pf2}, we get
\begin{equation}
\sum_{k=1}^K\left(f^\prime\hspace{-1pt}\left(e_k\right)e_k + f\hspace{-1pt}\left(g\hspace{-1pt}\left(f^\prime\hspace{-1pt}\left(e_k\right)\right)\right) - f^\prime\hspace{-1pt}\left(e_k\right)g\hspace{-1pt}\left(f^\prime\hspace{-1pt}\left(e_k\right)\right)\right).
\label{opt23}
\end{equation}

Since we have $g\hspace{-1pt}\left(f^\prime\hspace{-1pt}\left(e_k\right)\right) = f^\prime\hspace{-1pt}\left(g\hspace{-1pt}\left(e_k\right)\right) = e_k$,\footnote{In fact, it can be shown that there is not a scalar function $g(x)$ such that it is also left inverse of  $f^\prime(x)$. However, the relation $g\hspace{-1pt}\left(f^\prime\hspace{-1pt}\left(e_k\right)\right) = f^\prime\hspace{-1pt}\left(g\hspace{-1pt}\left(e_k\right)\right) = e_k$ can be obtained by defining a non-scalar function \cite[Rem. 3]{demir2020lsfp}. Since it is not required to explicitly know $g(x)$, we can assume $g\hspace{-1pt}\left(f^\prime\hspace{-1pt}\left(e_k\right)\right) = f^\prime\hspace{-1pt}\left(g\hspace{-1pt}\left(e_k\right)\right) = e_k$ without loss of generality.} and owing to the fact that the expression of the MSE $e_k$ does not rely on the chosen utility, the update of the auxiliary variables $v_{k}^{(n)}$ and $e_k^{(n)}$ are the same as in the sum-SE maximization as mentioned earlier, hence \eqref{opt23} gives
\begin{equation}
\sum_{k=1}^Kf\hspace{-1pt}\left(e_k\right) = \sum_{k=1}^K-\textrm{ln}\left(-\textrm{ln}\left(e_k\right)\right) = -\sum_{k=1}^K\textrm{ln}\left(\textrm{ln}\left(1 + \textrm{SINR}_k\right)\right)
\label{opt24}
\end{equation}

As a result, minimizing the objective in \eqref{pf2} is the same as maximizing the objective in \eqref{pf}. Modifying the update of the auxiliary variable $\omega_k^{(n+1)}$ in \eqref{eq:Updated weight}, which has been derived from \cite[The.~2]{shi2011iteratively}, we can employ Algorithm~\ref{alg1} together with the ADMM to obtain a local optimum solution to the PF maximization problem. The sum-SE and PF maximization problems are solved to produce training data for the machine learning algorithm. To manage a large setup, we need reasonably low computational complexity, which is obtained thanks to the derived closed-form updates of the WMMSE-ADMM algorithm. Other utility functions may be used for the setup at hand, for example, it is straightforward to include weights in the same optimization objectives. However, the chosen sum-SE and PF maximization schemes already cover practical choices for power allocation by mobile networks that are based on SE maximization.

\section{Distributed DNN-Based Power Allocation}\label{secDist}

Each of the optimization algorithms described in the previous section maximizes a network-wide performance objective using network-wide information, thus it must be implemented at the CPU. A main objective of this paper is to develop distributed algorithms that can be implemented separately at each AP, based on locally available information, while still aiming at maximizing a network-wide performance objective.

In this section, we develop learning-based solutions in the form of distributed DNNs to learn the optimal power coefficients with significantly lower fronthaul requirements compared to either centralized deep learning solutions or network-wide optimization problems, both required to be solved at the CPU. Different from the traditional optimization approach for solving \eqref{sumse} or \eqref{pf} that require knowledge of $\{\mathbf{a}_{k}\}$ and $\{\mathbf{B}_{ki}\}$, we demonstrate that LSF information can be utilized for computing an approximation of the optimal powers directly. We stress that, in the proposed solution, each AP has a unique DNN that is trained based on the local propagation environment of that AP and its relative location with respect to the other APs. This stands in contrast to the existing fractional power allocation algorithms that apply the same heuristic at every AP \cite{interdonato2019scalability,demir2021foundations}.
Fig.~\ref{model1} shows the cell-free network architecture. The implementation of the distributed DNN models can be done at the AP level on dedicated hardware optimized for machine learning, providing more efficiency compared to general-purpose processors \cite{browncloud}.

\subsection{Fully Distributed DNN}\label{fullydist}

This section presents the proposed fully distributed per-AP DNN-based power allocation model which relies only on local statistical information about the channels between the local AP and the different UEs, which can be easily obtained at an AP. For a given AP $l$, we employ a fully-connected feedforward DNN that takes as input the UEs’ LSF coefficients $\{\beta_{kl}: \forall k\}$ to perform power allocation as these parameters already capture the main features of the propagation channels and interference in the network, and can be easily measured in practice based on the received signal strength.\footnote{For example, one can compute the sample variance over the channel realizations obtained in different coherence blocks and also use the multiple antennas to obtain additional observations \cite{bjornson2017book}.} Hence, for given locations of the APs, we attempt to learn the unknown mapping between the locally available coefficients $\{\beta_{kl}: \forall k\}$ and the optimal square-roots of the transmit powers $\{\mu_{kl}^{\star}: \forall k\}$.\footnote{Since we are using fewer inputs than the centralized optimization algorithms, there might not be a unique mapping. Even with infinite training time, a perfect mapping might not be achieved. However, we can still fill in parts of the missing information by learning the local propagation environment and how likely different solutions are to appear, i.e., how users are distributed.} This alleviates the necessity to exchange LSF coefficients among the APs, in contrast to a centralized implementation, allowing for a scalable network operation.

\subsubsection{Input Data Preparation}

To simplify the training, it is essential to pre-process the input data based on the available domain knowledge, so the training can focus on learning the unknown aspects.
Hence, instead of having the LSF coefficients as a raw input to the DNN, a heuristic power allocation scheme based on the LSF coefficients is adopted to obtain the inputs. We have discovered numerically that the proposed heuristic input provides better dynamic range and distribution, resulting in an improved performance of the distributed DNN. The coefficients obtained from the heuristic, for a given sample, are computed in a similar manner to \cite{interdonato2019scalability} as
\begin{equation}
\rho_{kl}^{\prime} = \sqrt{P_{\textrm{max}}^{\textrm{dl}}}\frac{\left(\beta_{kl}\right)^v}{\sum_{i = 1}^K\left(\beta_{il}\right)^v}, \quad k = 1,\hdots, K,
\label{input}
\end{equation}
where $v$ is a constant  exponent that reshapes the LSF coefficients. Note that the ratio between the LSF coefficients can be directly obtained from the heuristic in \eqref{input}. This ratio is the main factor upon which the power allocation is determined. Accordingly, the heuristic input provides better dynamic range and distribution as stated, without loss of information. After which, standardization of the dataset is done by utilizing a \textit{robust} scaler to the logarithm (dB scale) of the heuristic coefficients $\{\rho_{kl}^{\prime}: \forall k\}$. It scales the data using the range between the first and third quartiles, i.e., the interquartile range. This, in turn, reduces the effect of outliers in the dataset that exist due to the large differences in the LSF coefficients from the different UEs to a given AP, particularly seen for relatively large coverage areas.

\subsubsection{Structure of the Proposed DNN}

The layout of the fully distributed DNN for each AP is shown in Table \ref{fullydisttable}. Note that the DNN not only attempts to approximate the normalized power coefficients, but also approximates the total power transmitted by the AP, that is (for AP $l$): $\sum_{k = 1}^K\mu_{kl}^2$. We observed that adding this constraint, in the form of an extra output, to the network improves the accuracy of the predicted powers. The reason is that the sum of the allocated powers by an AP obtained from the WMMSE algorithm is not necessarily constant and equal to the maximum power limit $P_{\max}^{dl}$, even though $P_{\max}^{dl}$ is the same for all APs. In fact, for most of the simulated UE locations, the total allocated power by a given AP is less than $P_{\max}^{dl}$, i.e., the WMMSE algorithm chooses to cut down on the power from some of the APs in its attempt to maximize the presented optimization objective. This results in the output layer size being $K+1$ instead of $K$. Further details about the choice of activations are provided in Section \ref{activations}.

\begin{table}
\begin{center}
\caption{\centering Layout of fully distributed DNN for an AP.
Parameters to be trained: 5,557}
\begin{tabular}{ c|c|c|c  }
 & Size & Parameters & Activation Function\\
\hline
Input & $K$ & - & - \\ 
Layer 1 (Dense) & $32$ & $672$ & linear \\
Layer 2 (Dense) & $64$ & $2112$ & tanh \\
Layer 3 (Dense) & $32$ & $2080$ & tanh \\
Layer 4 (Dense) & $K + 1$ & $693$ & relu \\
\end{tabular}
\label{fullydisttable}
\end{center}
\end{table}

\subsection{Distributed DNN with Side Information}

To improve the performance of the distributed learning-based solution, another distributed DNN model is proposed in this section. This DNN model utilizes side information that can be computed without the need to communicate information to the APs through the fronthaul links, in addition to the input described in Section \ref{fullydist}.

\subsubsection{Side Information}

A limiting factor to the distributed solution is the lack of information about the inter-relations between the LSF coefficients of a given UE to all APs in the network. Basic relationships can be implicitly learned from the data, but not everything. A typical basic relationship is the proportion of the UEs' average channel gains from one AP, to how large the contribution of this AP is to the SE of a given UE. This contribution is determined by the maximization objective that the NNs are being trained to approximate. However, the SE expression of this UE is not only affected by the average channel gains of all UEs to this single AP, but also by the channel gains of the UEs to the other APs. The local LSF coefficients at a given AP cannot fill in this information.
For example, a UE with a given LSF can be located in different angular directions from the considered AP, whereof some will lead to it being close to other APs and some will not, thereby impacting how essential the power transmitted by the considered AP will be for the UE.
The proposed extra input to the DNN partly fills in this missing information by providing a ratio of the LSF coefficients from the AP, to the LSF coefficients from all APs in the network. There is one such ratio per UE, which can be formulated as
\begin{equation}
\rho_{kl}^{\dprime} = \sqrt{P_{\textrm{max}}^{\textrm{dl}}}\frac{\left(\beta_{kl}\right)^v}{\sum_{l^{\prime} = 1}^L\left(\beta_{kl^{\prime}}\right)^v}, \quad k = 1,\hdots, K.
\label{si}
\end{equation}

A \textit{robust} scaler is utilized to scale this additional input to the DNN after converting it to dB scale, as previously described in Section \ref{fullydist}. To acquire this side information at each AP without the need for fronthaul signaling, the following procedure can be adopted. First, the denominator of \eqref{si} is estimated at UE $k$ based on the received signal strength from all APs to that particular UE. Then, each UE broadcasts this single combined value to all APs in the network using standard control signaling channels. Since each AP already has the information about the LSF coefficients of the UEs to itself (numerator of \eqref{si}) as described earlier, the side information is computed at no extra cost in terms of fronthaul signaling.

\subsubsection{Structure of the Proposed DNN}

The layout of the distributed DNN with side information is shown in Table \ref{distsitable}. The input layer is composed of a vector of size $2K$ in this case. That is, for every UE there exist two inputs to each AP DNN, which are as presented in \eqref{input} and \eqref{si}. As can be seen, to utilize the extra information available to the DNN, a more complicated structure for the DNN is adopted in comparison to the fully distributed case. The motivations behind the choice of activations are provided in Section \ref{activations}.

\begin{table}
\begin{center}
\caption{\centering Layout of distributed DNN with side information. Parameters to be trained: 21,973}
\begin{tabular}{ c|c|c|c  }
 & Size & Parameters & Activation Function\\
\hline
Input & $2K$ & - & - \\ 
Layer 1 (Dense) & $64$ & $2624$ & linear \\
Layer 2 (Dense) & $128$ & $8320$ & elu \\
Layer 3 (Dense) & $64$ & $8256$ & tanh \\
Layer 4 (Dense) & $32$ & $2080$ & tanh \\
Layer 5 (Dense) & $K + 1$ & $693$ & relu \\
\end{tabular}
\label{distsitable}
\end{center}
\end{table}

\section{Clustered DNN-Based Power Allocation}\label{secClu}

\begin{figure}
\centering
\setlength{\abovecaptionskip}{0.3cm plus 0pt minus 0pt}
\includegraphics[scale=0.67]{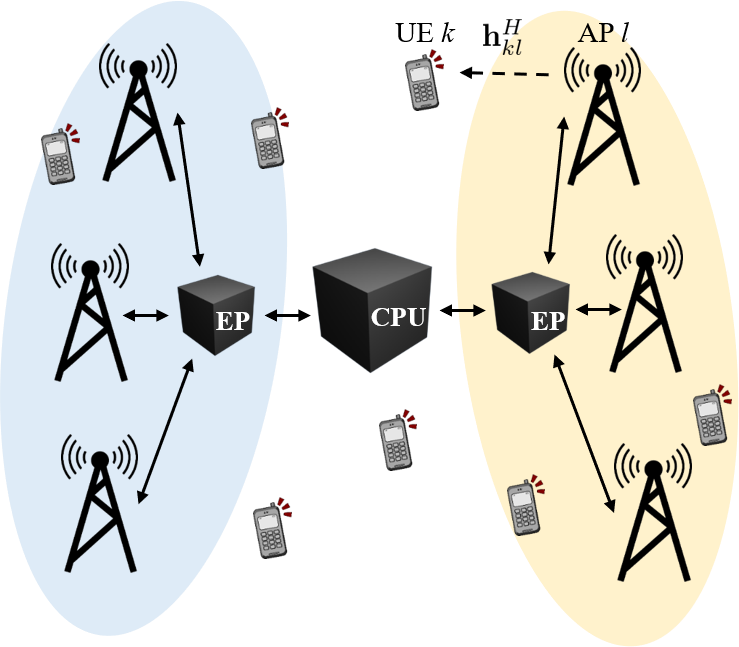}
\caption{Alternative cell-free network architecture with clustered APs connected to the CPU via EPs. The solid lines represent fronthaul/midhaul links.}
\label{model2}
\end{figure}

Large cell-free networks will likely be hierarchical, where disjoint clusters of APs are connected to separate edge processors (EPs) \cite{interdonato2019scalability}, which in turn are interconnected with the CPU (which might be a physical or logical entity). This network architecture is illustrated in Fig.~\ref{model2}. 
In this section, we propose an alternative learning-based solution to the power allocation problem that exploits this structure by having a joint DNN for power allocation in each cluster.
There is a connection between the considered architecture and the hybrid cloud radio access network (H-CRAN) presented in \cite{sriram2019joint}, which exemplifies that cell-free massive MIMO can be deployed on top of that practical architecture.

The proposed clustered DNN models can be implemented at the EPs on dedicated hardware optimized for machine learning, and accordingly the LSF coefficients of the APs belonging to a given cluster need to be communicated to the EP through fronthaul links. However, no LSF coefficients are required to be shared between the different clusters. This represents an alternative solution to the problem of lack of information about the inter-relations between the LSF coefficients of a given UE to the APs in the network that was mentioned earlier. The clustered architecture is preferred over a star topology in terms of decreasing the length of the fiber connections and deployment cost, especially when considering wireless fronthaul links from the APs of each cluster to the nearby EPs \cite{sriram2019joint,habibi2019comprehensive}. Moreover, edge computing can offer an ultra-low latency environment with high data rate, enabling computationally intensive and delay-sensitive applications, as the power allocation problem, to be executed in close proximity to the UEs \cite{habibi2019comprehensive}. The clustered DNN-based power allocation provides a trade-off between the achievable SE in the network and the fronthaul/midhaul requirements. Again, since the clustered DNNs make use of the network-wide sum-SE and PF solutions in the training phase, each EP employs a DNN that implicitly makes use of the entire network structure and propagation environment.

The layout of the clustered DNN is shown in Table \ref{clustered}. The number of APs per cluster is denoted by $c$. It is worth mentioning that the structure of the DNN is dependent on the chosen cluster size. The presented layout is chosen based on numerical evaluations for $c = 3$, however, it can be suitable for different cluster sizes such that $cK < 128$. The input to the DNN is formed by employing a \textit{robust} scaler, described in Section \ref{fullydist}, to the dB scale of the LSF coefficients from all UEs to the APs belonging to the cluster. The output of the DNN of a cluster represents the normalized power coefficients from all APs in the cluster to all UEs in the network, alongside the predicted total power for each AP in the cluster.

\begin{table}
\begin{center}
\caption{\centering Layout of clustered DNN.
Parameters to be trained: 246,207}
\begin{tabular}{ c|c|c|c  }
 & Size & Parameters & Activation Function\\
\hline
Input & $cK$ & - & - \\ 
Layer 1 (Dense) & $128$ & $7808$ & linear \\
Layer 2 (Dense) & $512$ & $66048$ & elu \\
Layer 3 (Dense) & $256$ & $131328$ & tanh \\
Layer 4 (Dense) & $128$ & $32896$ & tanh \\
Layer 5 (Dense) & $c(K + 1)$ & $8127$ & relu \\
\end{tabular}
\label{clustered}
\end{center}
\end{table}

\section{Training, Activations, and Complexity}\label{complexity}

This section focuses on the training procedure and the motivation for the structure of the DNN models presented earlier. In addition, we provide a brief on the computational complexity and run-time requirements of the different conventional and learning-based algorithms, to solve the power allocation problem, that are employed within this work.

\subsection{DNN Training}

For training the DNNs, first a set containing a large number of training pairs $\left\{\mathit{f}\hspace{-2pt}\left(\beta_{kl}\right), \mu_{kl}^{\star}: \forall k, l\right\}$ is generated. The training of all proposed DNN models was done on two different datasets: one with orthogonal pilots ($\tau_p = K$) and one with non-orthogonal pilots ($\tau_p < K$). The datasets were generated by solving the optimization problems using the WMMSE-ADMM algorithm described in Section~\ref{optimization}. We define the vector $\boldsymbol{\bar{\mu}}_{l}^{\star} = [\bar{\mu}_{1l}^{\star}, \hdots, \bar{\mu}_{Kl}^{\star}, \,\Omega_l]^T$ as the concatenation of the optimal normalized powers from AP $l$ to all UEs in the network alongside the total power allocated by the AP, represented as a ratio of the total power budget such that $\Omega_l = \frac{\sum_{k = 1}^{K}\mu_{kl}^2}{P_{max}^{dl}}$. In the distributed cases, the DNN for AP $l$ is trained to minimize the following loss:
\begin{equation}
\textrm{Loss} = \norm{\boldsymbol{\bar{\mu}}_{l}^{\star} - \boldsymbol{\bar{\mu}}_{l}^{\textrm{DNN}}}^2.
\label{loss1}
\end{equation}

As for the clustered DNN, let the corresponding vector of optimal normalized powers from all APs and total allocated power ratio in a given cluster to all the UEs in the network be denoted by $\boldsymbol{\eta}^{\star} = [\boldsymbol{\bar{\mu}}_{l}^{\star}, \hdots, \boldsymbol{\bar{\mu}}_{l+c-1}^{\star}]^T$. Accordingly, the loss function for the DNN of a cluster can be written as
\begin{equation}
\textrm{Loss} = \norm{\boldsymbol{\eta}^{\star} - \boldsymbol{\eta}^{\textrm{DNN}}}^2.
\label{loss2}
\end{equation}

The loss functions described in \eqref{loss1} and \eqref{loss2} are averaged over the training samples and the DNNs aim at minimizing these loss functions for the distributed and clustered models, respectively. The training process contains updating of weights between neurons as well as the bias in each layer. The Adam optimizer is employed and the learning rate is set to $0.001$ for the initial epochs to ensure convergence, after which it is decreased by a factor of $10$ to fine tune the weights update. The batch size and number of epochs are chosen by a trial-and-error method.

\subsection{Choice of Activation Functions}\label{activations}

While we propose specific DNN structures, it is worth mentioning that finding the best DNN structure and activation functions can also be seen as optimization problems on their own, requiring further research. Referring to the literature \cite{goodfellow2016deep,zhao2020power}, we tried several structures and activations to find the configurations providing the least MSE for our problem. It can be noticed that the proposed DNNs follow, to a certain extent, similar structures. \textit{Linear} and \textit{elu} activations are used in the ``expansion"  layer(s), where the layer size is increased from one layer to the next, to emphasize the differences between the inputs. This is particularly important as the LSF coefficients from each AP to the randomly dropped UEs within a large coverage area vary greatly forming a wide range of possible values, whereas a large portion of the LSF coefficients are relatively small compared to this range of values. Afterwards, while decreasing the layer sizes gradually towards the desired output dimension, \textit{tanh} activations are used to balance the resulting gradients from \textit{linear} activations and avoid the exploding gradients problem, providing more stability for the DNN models.

\subsection{Complexity Analysis}

By analyzing the update steps of the WMMSE-ADMM algorithm, it can be noticed that the $L \times L$ matrix inverse in \cite[Eq.~(48)]{chakraborty2020efficient} has the greatest computational complexity in terms of big-O. This operation needs to be performed only once for each UE per each problem, since the same inverse can be used in subsequent iterations. Accordingly, the computational complexity of the WMMSE-ADMM algorithm is $\mathcal{O}\left(L^3K\right)$ \cite{chakraborty2020efficient}.

The computational complexity of the learning-based approach is mostly in the data generation phase for training the DNN, where the actual optimization problems in \eqref{sumse} or \eqref{pf} need to be solved repeatedly to produce the training dataset. On the other hand, for a DNN with $T$ layers such that a layer has $T_i$ neurons, the required number of real multiplications and additions is each $T_iT_{i - 1}$, at layer $i$, for $i = 1, \hdots, T$. Moreover, $\sum_{i = 1}^TT_i$ activation functions need to be evaluated in total.

In the following, the fully distributed DNN will be referred to as ``DDNN", the distributed DNN with side information as ``DDNN-SI" and the clustered DNN as ``CDNN". The estimated run-time of the DNN models is significantly lower than the conventional approach of solving the optimization problem using the WMMSE-ADMM algorithm. Nevertheless, it is worth stating that it is difficult to make a fair comparison in terms of processing time on a practical implementation. An important reason is that the implementations will be based on different hardware architectures. The WMMSE-ADMM algorithm is well suited for execution on a classical (multicore) CPU architecture, whereas the DNN will likely run on a hardware architecture optimized for machine learning, e.g., a neural processing unit (NPU) \cite{zhao2020power}. That said, it is still revealing to see the huge run-time differences when both are executed on the same hardware. In Table \ref{runtime}, we recorded the average run-time, for 100 samples, of the WMMSE-ADMM algorithm and the DNN models for the processing schemes and optimization objectives presented earlier. It can be seen that the DNN models satisfy real-time processing constraints that mainly rely on scheduling decisions in the network, which can be in the order of 1\,ms as previously stated. Note that the recorded run-times for the learning-based solutions represent the total time taken by all the distributed/clustered DNNs implemented to support the entire cell-free network, i.e., the computational time for each DNN is less than 1\,ms. It is clear that the CDNN provides the least (total) computational complexity by facilitating a common DNN for the APs in each cluster. We use the same platform, a 4 core Intel(R) Core i5-10310U CPU with 1.7 GHz base frequency and 4.4 GHz max turbo frequency. The programs are all written in Python 3.8.

\begin{table}
\begin{center}
\caption{\centering Computational time in milliseconds for the proposed DNN models in comparison to the WMMSE-ADMM algorithm}
\begin{tabular}{ |c|c|c|c|c|  }
\hline
\multirow{2}{*}{Algorithm} & \multicolumn{2}{c|}{Sum-SE} & \multicolumn{2}{c|}{PF} \\
& MR & RZF & MR & RZF \\
\hline
ADMM & $88.6$ & $131.7$ & $124.8$ & $163.5$\\
DDNN & $8.8$ & $9.2$ & $8.6$ & $9.0$\\
DDNN-SI & $9.7$ & $9.8$ & $9.5$ & $9.7$\\
CDNN & $3.2$ & $3.2$ & $3.2$ & $3.2$\\
\hline
\end{tabular}
\label{runtime}
\end{center}
\end{table}

\section{Performance Evaluation}\label{numerical}

\subsection{Simulation Setup}

We consider a cell-free network comprised of $L = 16$ APs deployed in an area of $1000$\,m $\times$ $1000$\,m. The number of antennas per AP is $N = 4$. A wrap around topology is assumed in order to simulate a large area without unnatural boundaries. Further, we assume that $K = 20$ UEs are randomly and uniformly dropped within the area of interest.\footnote{The uniform distribution does not impose a specific structure on the simulation scenario. It provides the least amount of information (most ambiguity) on the UE locations, since each location on the map has an equal probability of having a UE.} The results are averaged over a test dataset composed of $2000$ UE distributions, independent of the training dataset. We consider communication over a $20$\,MHz channel with a total receiver noise power of  $-94$\,dBm. The per-AP maximum DL transmit power is $P_{\textrm{max}}^{\textrm{dl}} = 1$\,W, whereas each UE has an UL power, in the pilot transmission phase, of $p_i = 100$\,mW. The coherence block length is chosen to be $\tau_c = 200$. Hereafter, the term ``orthogonal pilots" refers to the case where unique orthogonal pilots are assigned to all users, i.e., $\tau_p = K$, while ``non-orthogonal pilots" refers to the case $\tau_p = 10$, where the pilot assignment algorithm described in Section \ref{pAssign} is utilized. The pathloss model for generating the LSF coefficients is given by \cite{bjornson2019making,chakraborty2020efficient}\footnote{Note that shadow fading is not included in \eqref{pathloss} since the usual log-normal distribution would result in a channel model that is not spatially consistent \cite{sanguinetti2018deep}; that is, two UEs at almost the same location would experience  completely different pathlosses. A possible solution is to resort to channel models based on ray tracing or recorded measurements, but that is outside the scope of the considered distributed power allocation problem and left for future research.}
\begin{equation}
\beta_{kl} = -30.5 - 36.7 \textrm{log}_{10}\left(\frac{d_{kl}}{1\hspace{2pt}\textrm{m}}\right) \textrm{dB},
\label{pathloss}
\end{equation}
where $d_{kl}$ is the distance between UE $k$ and AP $l$.  The simulation parameters are summarized in Table \ref{params} and represent an urban microcell environment.

Training of the DNN models is performed on two datasets, for the orthogonal and non-orthogonal pilots cases, each having $350,000$ samples. The scaling parameter $v$ in \eqref{input} and \eqref{si} is chosen to be $0.6$ by a trial-and-error method. To handle variations in UE numbers, there exist two approaches for learning-based solutions in general: Either have different DNNs for different numbers of UEs, or train a single DNN to handle a flexible number of UEs. Due to the simplicity of the distributed/clustered models with a relatively small number of trainable parameters, we suggest that a collection of different DNNs can be trained for different UE numbers, and the final weights of all the DNNs stored at the AP or EP for distributed and clustered cases, respectively. The AP or EP then chooses the DNN corresponding to the current number of UEs in the system.

\begin{table}
\begin{center}
\caption{\centering Cell-free massive MIMO network simulation parameters}
\begin{tabular}{ |c|c| }
\hline
Area of interest (wrap around) & $1000$\,m $\times$ $1000$\,m \\ 
Bandwidth & $20$\,MHz \\
Number of APs & $L = 16$ \\
Number of UEs & $K = 20$ \\
Number of antennas per AP & $N = 4$ \\
Pathloss exponent & $\alpha = 3.76$ \\
Per-AP maximum DL transmit power & $P_{\textrm{max}}^{\textrm{dl}} = 1$\,W \\
UL transmit power & $p_i = 100$\,mW \\
UL/DL noise power & $-94$\,dBm \\
Coherence block length & $\tau_c = 200$\\
Pilot sequence length & $\tau_p = 10$ or $K$ \\
\hline
\end{tabular}
\label{params}
\end{center}
\end{table}

\subsection{Orthogonal vs. Non-Orthogonal Pilots}

\begin{figure}
\setlength{\abovecaptionskip}{0.3cm plus 0pt minus 0pt}
\begin{subfigure}{.5\textwidth}
\centering
\includegraphics[scale=0.23]{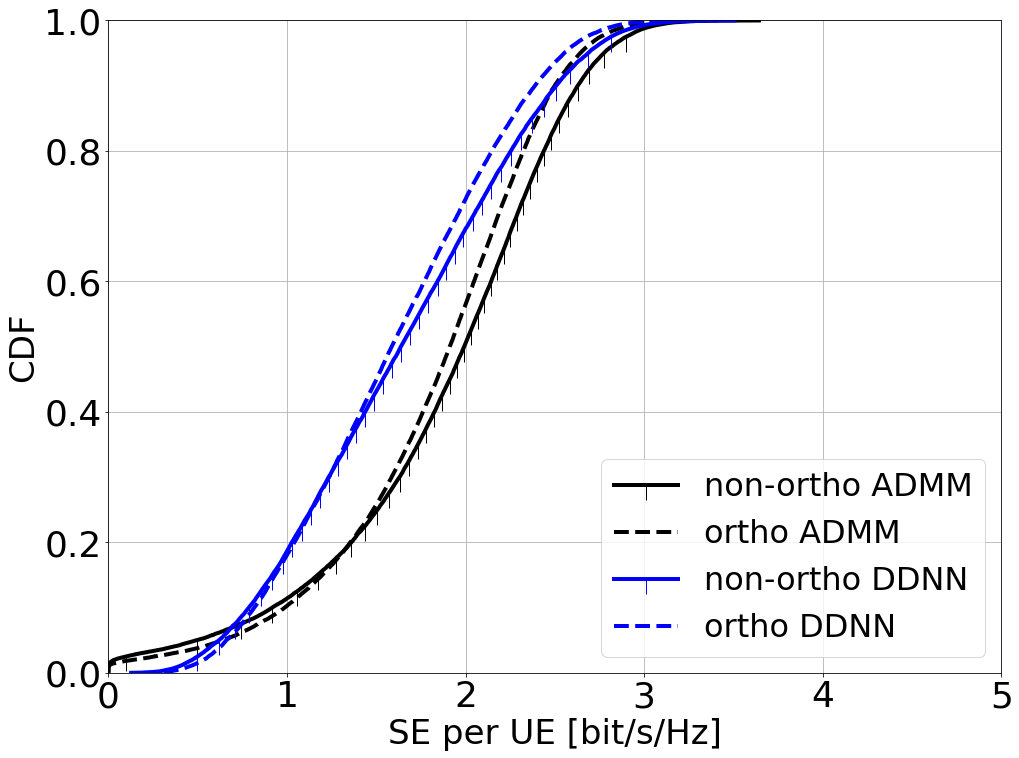}
\caption{MR DDNN comparison.}
\label{orthoMR}
\vspace{1em}
\end{subfigure}
\begin{subfigure}{.5\textwidth}
\centering
\includegraphics[scale=0.23]{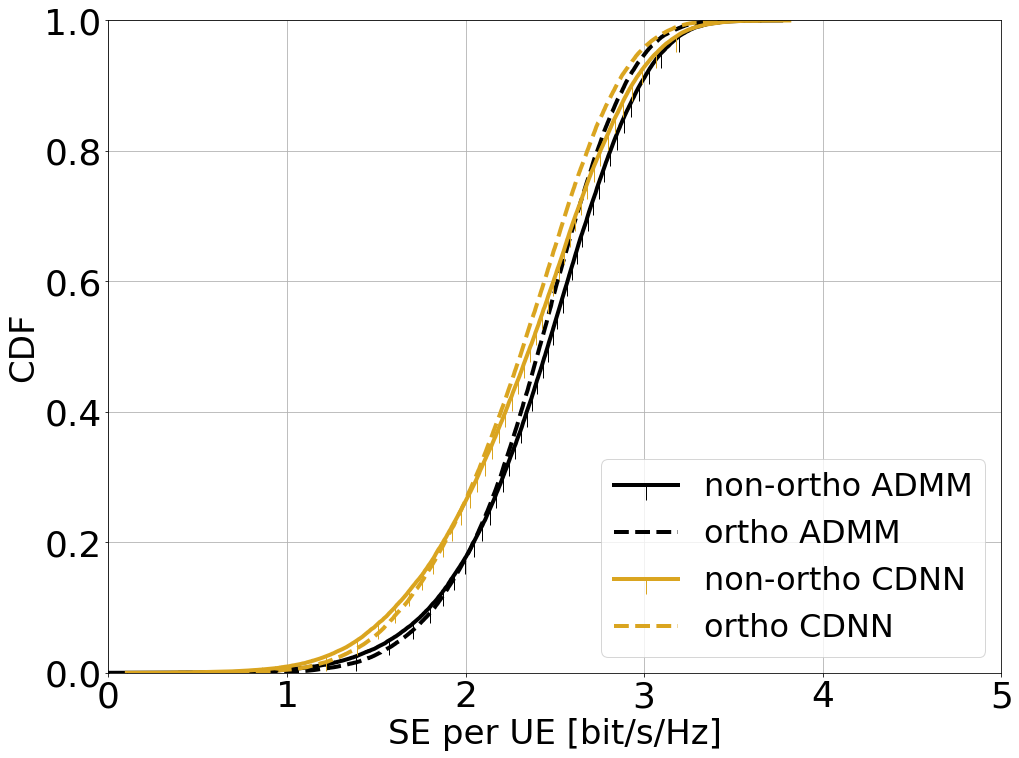}
\caption{RZF CDNN comparison.}
\label{orthoRZF}
\end{subfigure}
\caption{CDF of the DL SE per UE for orthogonal and non-orthogonal pilots with sum-SE maximization. Different precoding and power allocation are considered.}
\label{orthoFigure}
\end{figure}

In this subsection, we compare the results obtained from the two datasets with orthogonal and non-orthogonal pilots. We consider sum-SE maximization in  Fig.~\ref{orthoFigure}. In particular, Fig.~\ref{orthoMR} shows the cumulative distribution function (CDF) of the SE per UE with MR precoding for the DDNN model against the WMMSE-ADMM benchmark. The CDF is with respect to the random user locations. It can be seen that the capability of the DDNN to approximate the power coefficients is about the same with orthogonal and non-orthogonal pilots. There is a slight degradation in the performance of the less fortunate UEs for the case of non-orthogonal pilots due to the added interference in the channel estimation phase. On the other hand, UEs with better channel conditions experience an improvement with non-orthogonal pilots. The reason behind this is that the benefit of utilizing a larger portion of the coherence block for data transmission outweighs the degradation resulting from the added interference in the channel estimation phase. Fig.~\ref{orthoRZF} shows a similar outlook with RZF precoding and the CDNN approximation. However, the performance gain from using non-orthogonal pilots for UEs with better channel conditions is less than in the case of MR precoding since the channel estimation quality has a greater impact on performance for RZF precoding. All subsequent sections utilize the non-orthogonal pilots dataset, which is deemed the most realistic one.\footnote{We have observed that the proposed pilot assignment scheme provides a $5\,$-$10\, \%$ improvement in the total DL SE compared to random pilot assignment.}

\subsection{Sum-SE vs. PF Maximization}

\begin{figure}
\setlength{\abovecaptionskip}{0.3cm plus 0pt minus 0pt}
\begin{subfigure}{.5\textwidth}
\centering
\includegraphics[scale=0.23]{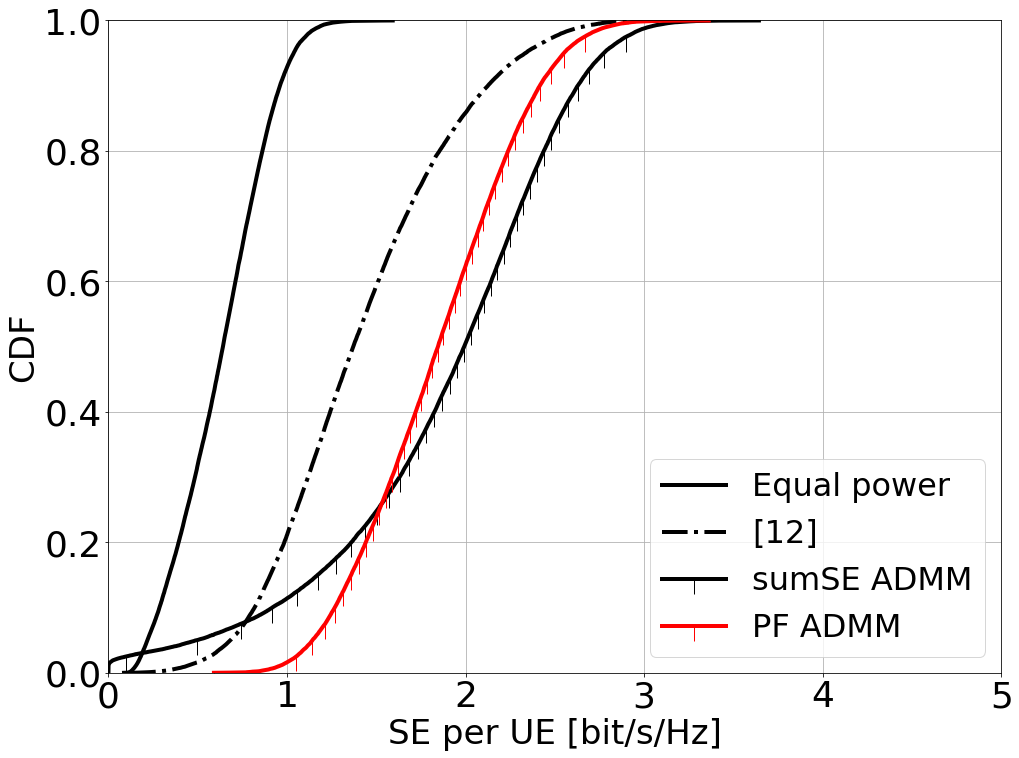}
\caption{MR.}
\label{MRcomparison}
\vspace{1em}
\end{subfigure}
\begin{subfigure}{.5\textwidth}
\centering
\includegraphics[scale=0.23]{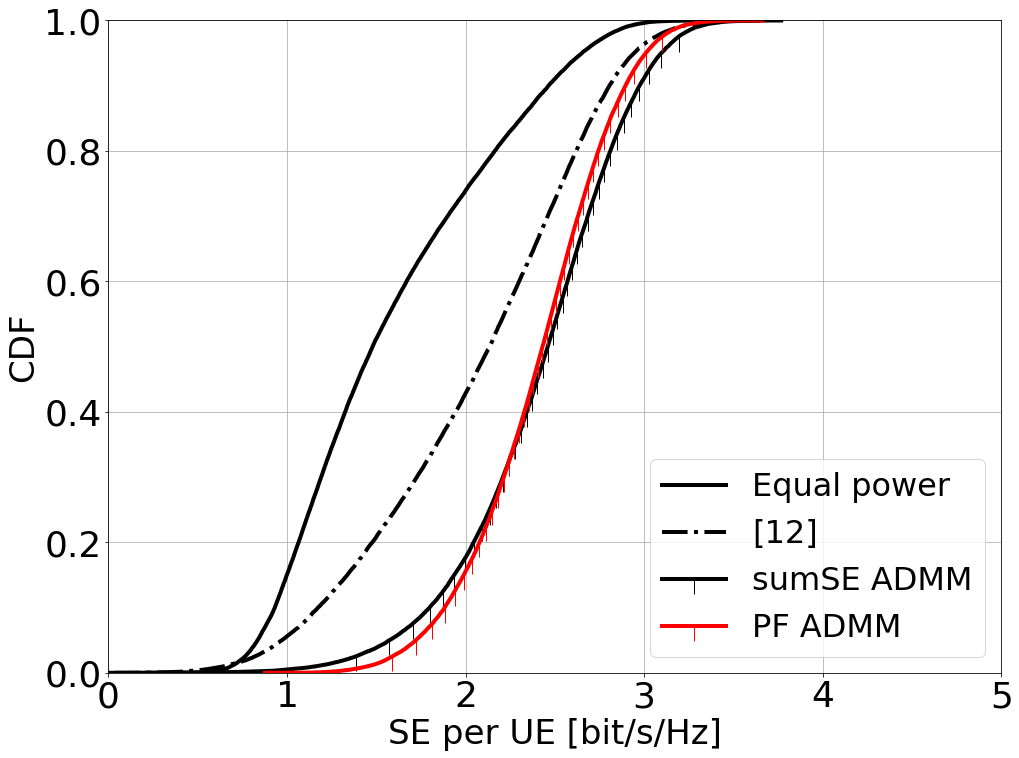}
\caption{RZF.}
\label{RZFcomparison}
\end{subfigure}
\caption{CDF of the DL SE per UE for sum-SE and PF maximization objectives.}
\label{sumSE_PF}
\end{figure}

In this subsection, we analyze the differences between the sum-SE and PF maximization objectives. For this purpose, Figs. \ref{MRcomparison} and \ref{RZFcomparison} plot the CDF of the SE per UE obtained by solving the sum-SE and PF optimization problems using the WMMSE-ADMM algorithm with MR and RZF precoding, respectively. The SEs obtained with equal power allocation as well as the heuristic scheme in \cite{interdonato2019scalability} are shown for comparison. It can be seen that with MR precoding, there is a clear difference between both schemes. While sum-SE solely focuses on maximizing the SE resulting in the possibility that less fortunate UEs being deprived of the network resources, PF attempts to find a reasonable trade-off between maximizing the SE and providing a good service for all the UEs within the coverage area.  With RZF, the performance difference between the two schemes is relatively small for the considered simulation setup. This is due to the capability of RZF to mitigate interference, and accordingly provide better SINRs even for the UEs with the worst channel conditions. This results in the sum-SE maximization problem yielding non-negligible power coefficients for all (or most) UEs.

\subsection{DDNN With Side Information Comparison}

\begin{figure}
\setlength{\abovecaptionskip}{0.3cm plus 0pt minus 0pt}
\begin{subfigure}{.5\textwidth}
\centering
\includegraphics[scale=0.23]{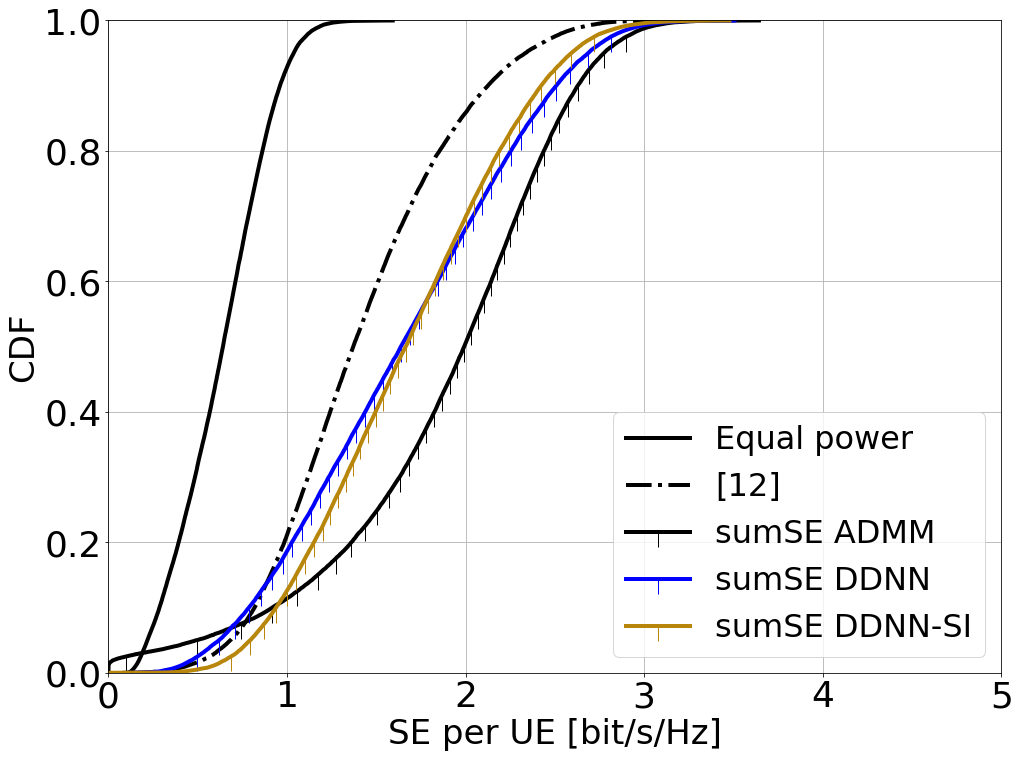}
\caption{Sum-SE.}
\label{MR_sumSE_DDNN}
\vspace{1em}
\end{subfigure}
\begin{subfigure}{.5\textwidth}
\centering
\includegraphics[scale=0.23]{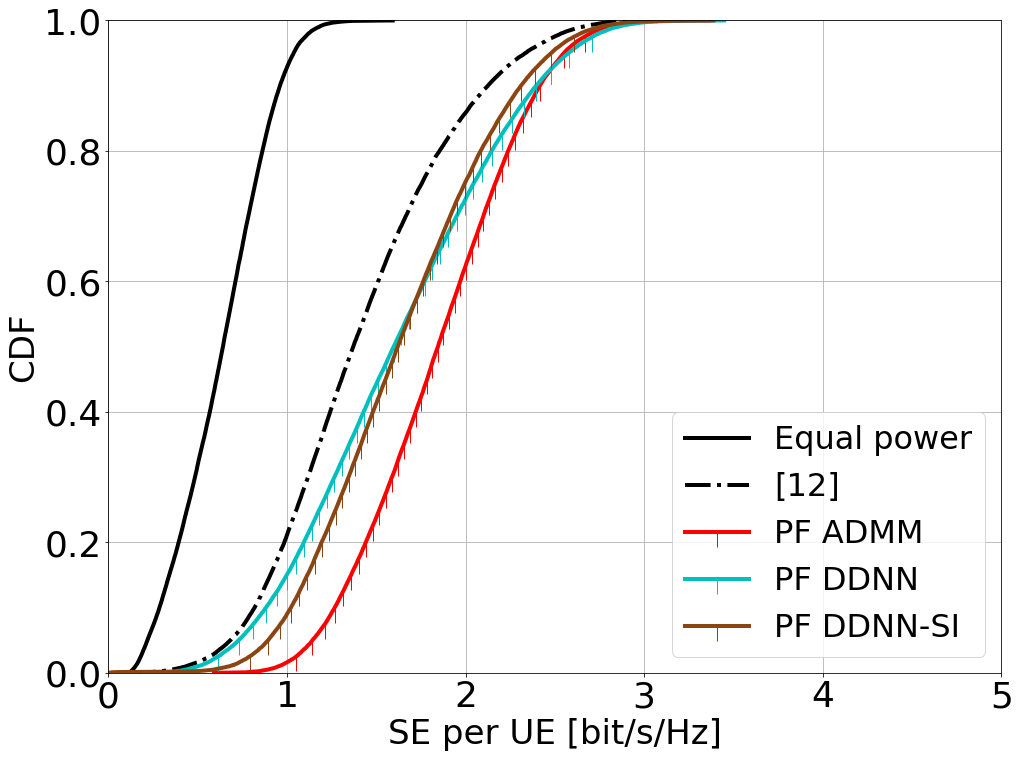}
\caption{PF.}
\label{MR_PF_DDNN}
\end{subfigure}
\caption{CDF of the DL SE per UE with MR.}
\label{MR_DDNN}
\end{figure}

\begin{figure}
\setlength{\abovecaptionskip}{0.3cm plus 0pt minus 0pt}
\begin{subfigure}{.5\textwidth}
\centering
\includegraphics[scale=0.23]{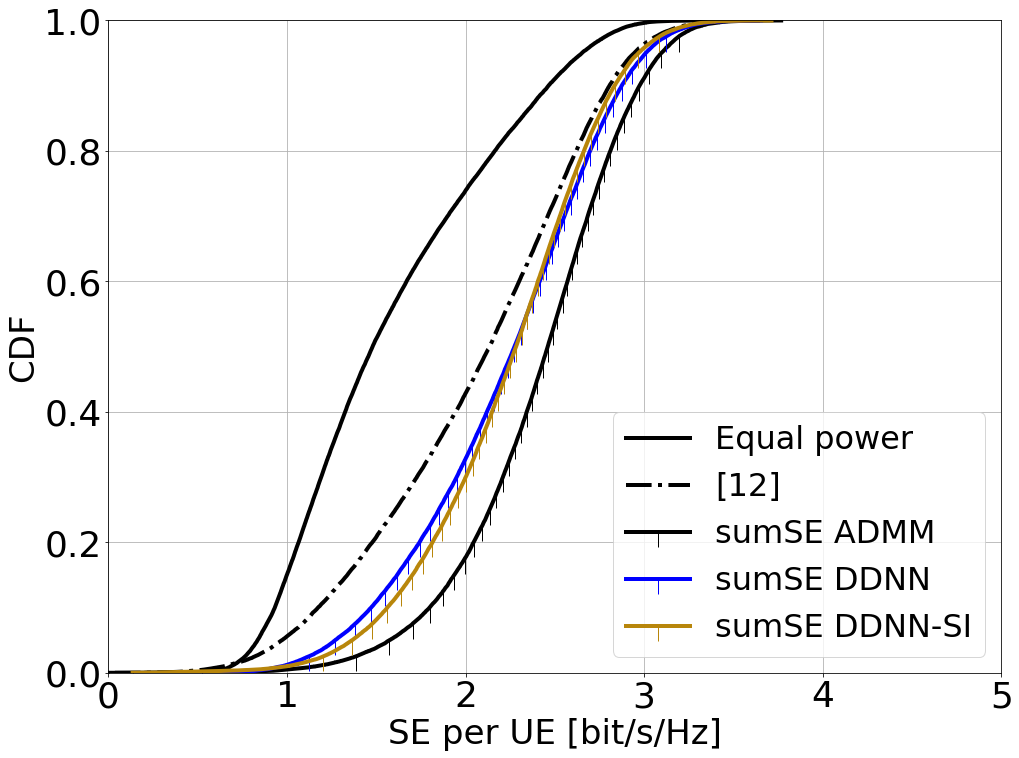}
\caption{Sum-SE.}
\label{RZF_sumSE_DDNN}
\vspace{1em}
\end{subfigure}
\begin{subfigure}{.5\textwidth}
\centering
\includegraphics[scale=0.23]{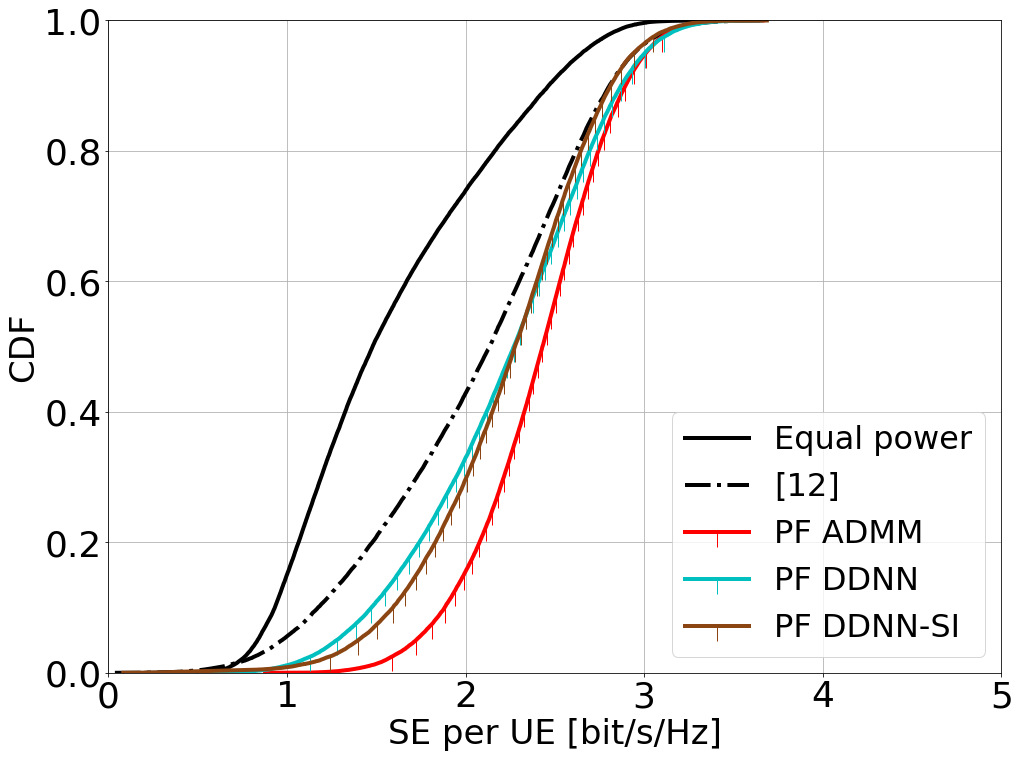}
\caption{PF.}
\label{RZF_PF_DDNN}
\end{subfigure}
\caption{CDF of the DL SE per UE with RZF.}
\label{RZF_DDNN}
\end{figure}

This subsection investigates the performance of the DDNN-SI model in comparison to the base case of the DDNN. The CDF of the SE per UE for sum-SE and PF maximization objectives with MR precoding is shown in Figs. \ref{MR_sumSE_DDNN} and \ref{MR_PF_DDNN}, respectively. In addition, equal power allocation as well as the heuristic scheme in \cite{interdonato2019scalability} are shown for comparison. The results show that the DDNN-SI provides better performance than the DDNN for the less fortunate UEs, i.e., UEs having low SE, at the expense of a slight degradation in performance for UEs with better channel conditions. This is a consequence of the extra input to the DDNN-SI of each AP which, as mentioned earlier, provides information about the inter-relations between the LSF coefficients of a given UE to all APs in the network. This information is more critical for the less fortunate UEs that are likely to be located in between several APs resulting in the LSF coefficients to those APs having comparable values. The DNN model thus utilizes this side information to help differentiate between the different angular directions that a given (single) LSF coefficient to one AP could possibly mean, and accordingly affects the output power coefficients of the DNN. A similar behaviour is seen in Fig.~\ref{RZF_DDNN} with RZF precoding. Moreover, the performance gap between the DNN models and the WMMSE-ADMM algorithm is smaller, showing that the capability of the DNN models to approximate the power coefficients is better than in case of MR precoding. Further, it is to be noted that, to the best of our knowledge, \cite{interdonato2019scalability} provides the highest SE among heuristic power allocation schemes. Figs. \ref{MR_DDNN} and \ref{RZF_DDNN} show that our distributed DNN models outperform this heuristic power allocation for all the considered precoding schemes and optimization objectives.

\subsection{CDNN Comparison}

\begin{figure}
\setlength{\abovecaptionskip}{0.3cm plus 0pt minus 0pt}
\begin{subfigure}{.5\textwidth}
\centering
\includegraphics[scale=0.23]{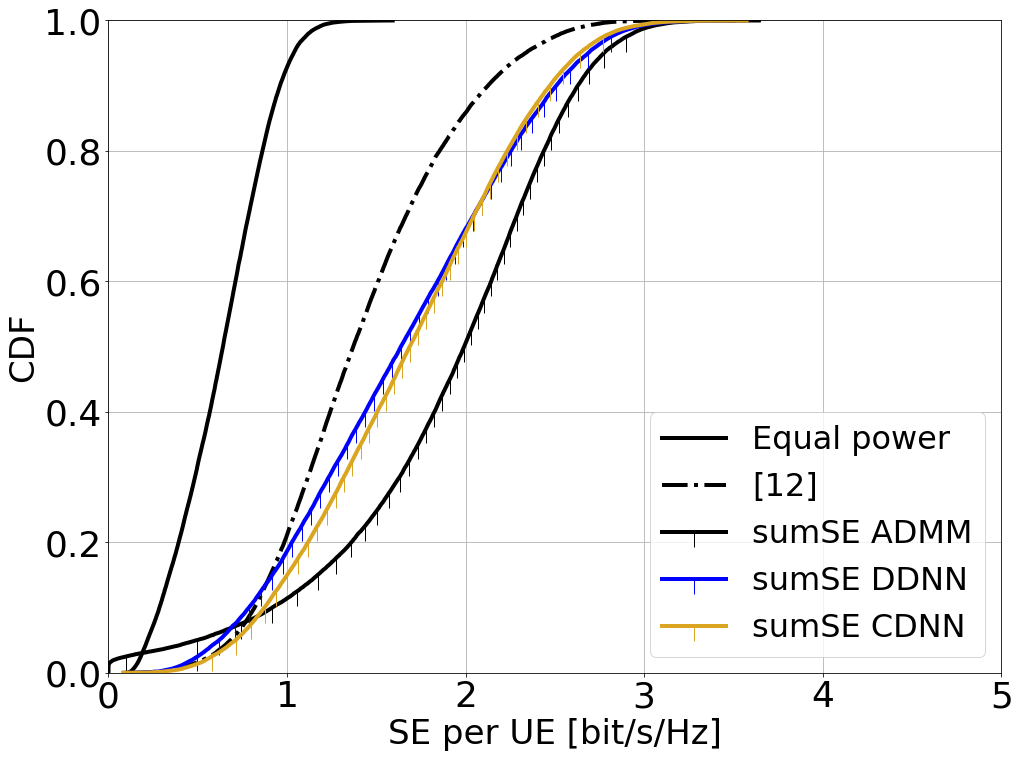}
\caption{Sum-SE.}
\label{MR_sumSE_CDNN}
\vspace{1em}
\end{subfigure}
\begin{subfigure}{.5\textwidth}
\centering
\includegraphics[scale=0.23]{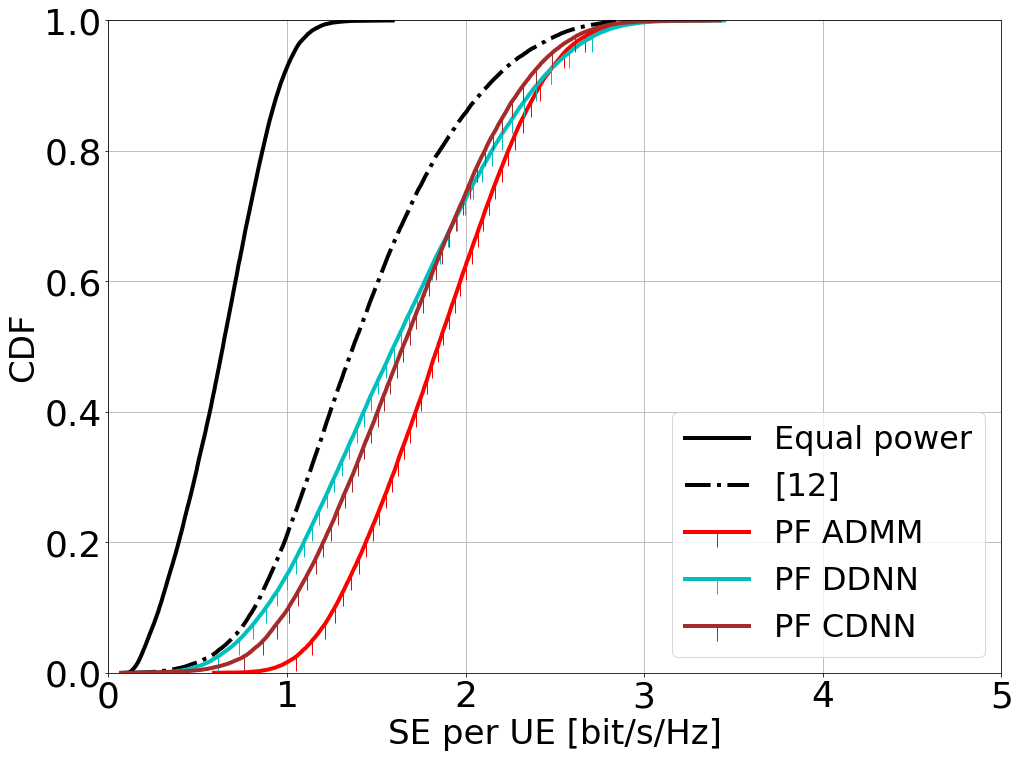}
\caption{PF.}
\label{MR_PF_CDNN}
\end{subfigure}
\caption{CDF of the DL SE per UE with MR.}
\label{MR_CDNN}
\end{figure}

\begin{figure}
\setlength{\abovecaptionskip}{0.3cm plus 0pt minus 0pt}
\begin{subfigure}{.5\textwidth}
\centering
\includegraphics[scale=0.23]{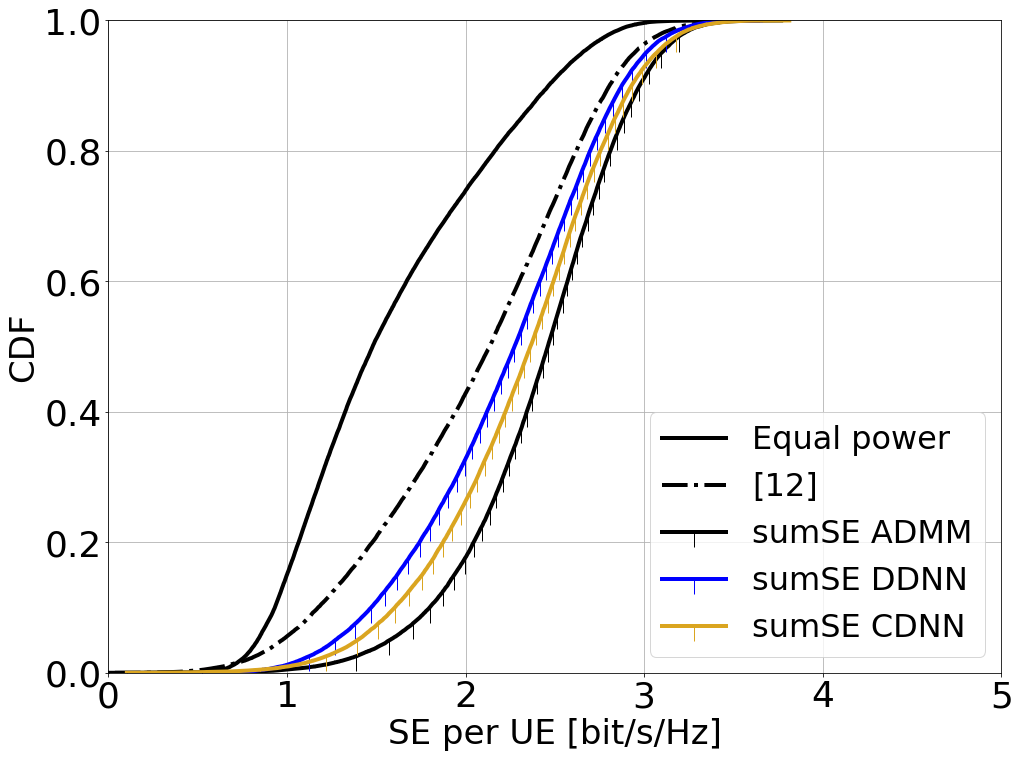}
\caption{Sum-SE.}
\label{RZF_sumSE_CDNN}
\vspace{1em}
\end{subfigure}
\begin{subfigure}{.5\textwidth}
\centering
\includegraphics[scale=0.23]{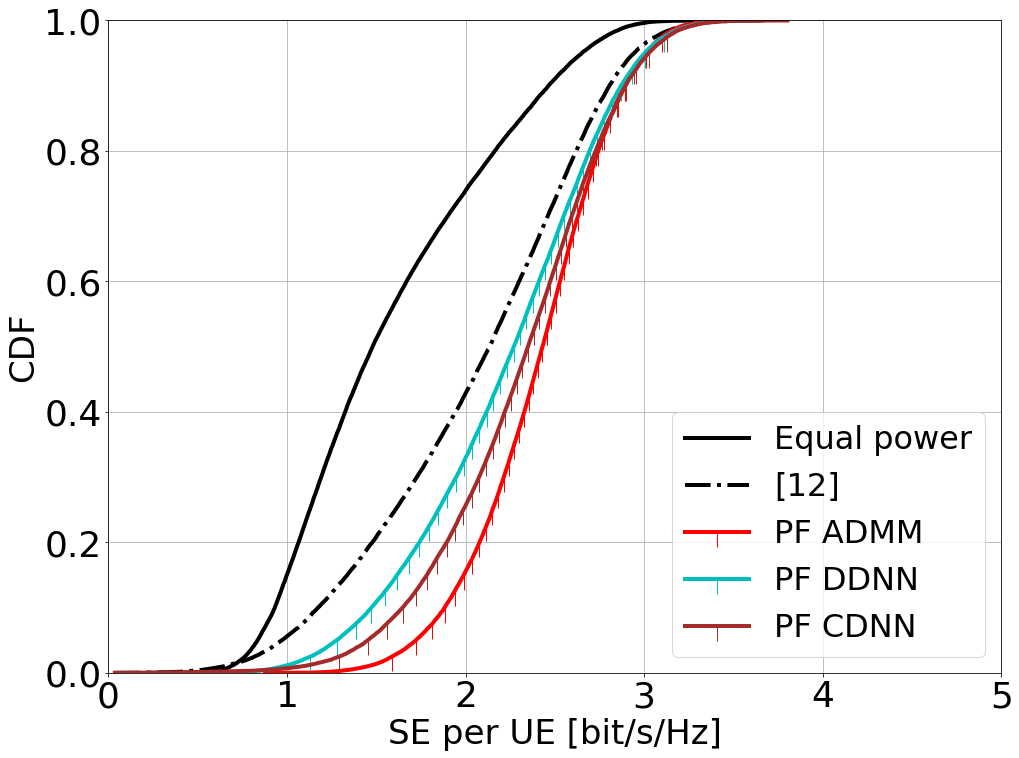}
\caption{PF.}
\label{RZF_PF_CDNN}
\end{subfigure}
\caption{CDF of the DL SE per UE with RZF.}
\label{RZF_CDNN}
\end{figure}

Another approach to solving the problem of lack of information about the inter-relations between the LSF coefficients is to use a clustered DNN implementation. This section evaluates the performance of the CDNN solution in comparison to the DDNN and heuristic power allocations. Fig.~\ref{MR_CDNN} shows the CDF of the SE per UE with MR precoding for the WMMSE-ADMM benchmark, the DDNN and CDNN solutions, as well as the heuristic and equal power allocations. It can be seen that the CDNN results in about the same performance as the DDNN-SI for the less fortunate UEs, but with a lower degradation for the UEs with better channel conditions compared to the DDNN. In Fig.~\ref{RZF_CDNN}, the performance of the CDNN with RZF precoding is investigated. It is seen that in case of RZF precoding, the CDNN provides an improvement over both the DDNN and DDNN-SI models. 

As suggested in the literature \cite{bashar2020exploiting,chakraborty2019centralized} and inferred from the results, with RZF precoding there is more room for improvement with learning-based approaches compared to MR precoding. The reason behind this is mainly that precoding schemes that aim at minimizing interference result in the power coefficients being more closely related to the LSF inputs. That is why combining inputs from several APs at an EP as in the CDNN implementation is more beneficial in this case. Other than that, the difference in terms of SE between the DDNN-SI and CDNN is rather small. These solutions are tailored for implementation on different physical network architectures on top of which cell-free massive MIMO is envisioned to be built in practice, so the choice of implementation will likely depend on the physical distribution of APs and network resources.

\subsection{Comparison of Total SE}

\begin{figure}
\centering
\setlength{\abovecaptionskip}{0.3cm plus 0pt minus 0pt}
\includegraphics[scale=0.23]{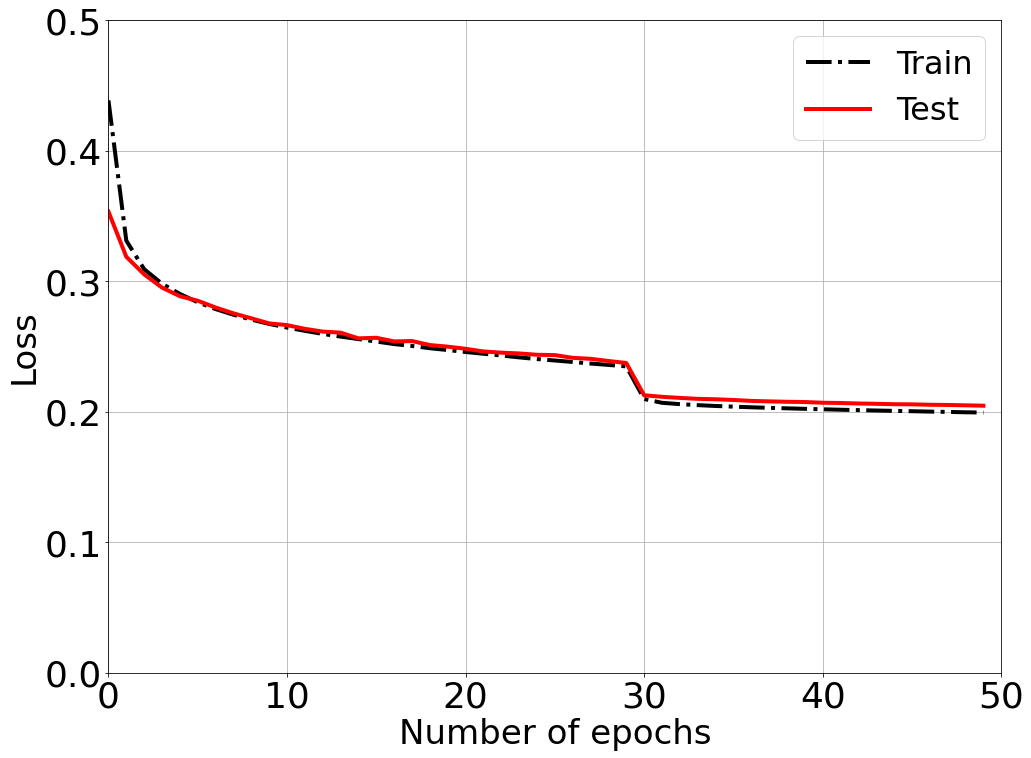}
\caption{Training curve for RZF CDNN.}
\label{training}
\end{figure}

This subsection investigates the total performance of the learning-based and conventional solutions presented earlier. The training curve for the CDNN with RZF precoding is shown in Fig.~\ref{training}. As can be seen in the figure, the loss function decreases when increasing the number of training epochs for the training and validation sets. Although the average loss for the power coefficients saturates at a relatively high value due to the smaller domain knowledge at the input of the DNN, this does not directly translate into a high performance gap compared to the ADMM. The reason behind this is that the SE of each UE is more influenced by the APs closest to it, and the capability of the DNNs to approximate the power coefficients of a given UE is dependent on the distance between the UE and the AP (which reflects in the magnitude of the LSF coefficient), i.e., the DNNs can better approximate the more important coefficients. This can be viewed as a natural inherent of the problem structure that is in favour of using a deep learning solution, and that even encourages a simplified distributed/clustered learning procedure without a significant degradation in the system performance.

\begin{figure}
\centering
\setlength{\abovecaptionskip}{0.3cm plus 0pt minus 0pt}
\includegraphics[scale=0.23]{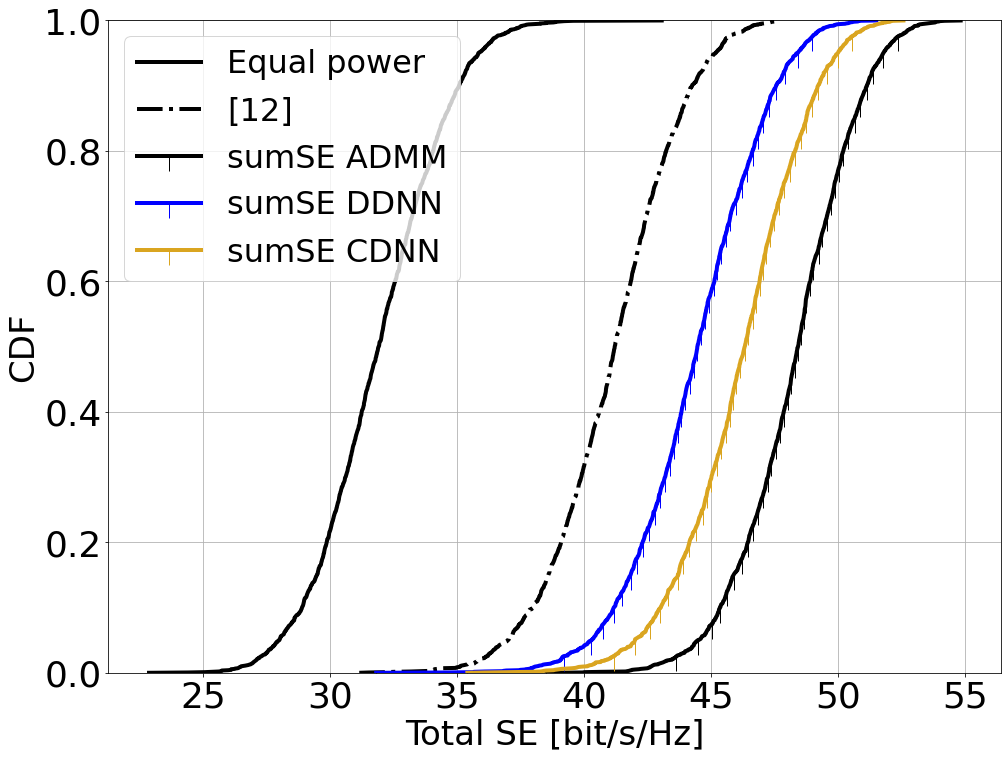}
\caption{CDF of the total DL SE for sum-SE maximization with RZF.}
\label{totalSE}
\end{figure}

To shed the light on the performance difference between the proposed deep learning solutions and practical LSF-based heuristic allocation strategies, as well as the gap to the WMMSE-ADMM benchmark, Fig.~\ref{totalSE} plots the CDF of the total DL SE for sum-SE maximization (which aims at maximizing the total SE) with RZF precoding. It can be seen that the proposed distributed and clustered DNNs provide a significant improvement over the heuristic in \cite{interdonato2019scalability}. Note that the DDNN-SI model achieves almost the same total DL SE as the CDNN model, and is thus omitted from the figure. Moreover, the average performance loss of the practical CDNN solution, providing reduced fronthaul and run-time requirements, compared to the centralized WMMSE-ADMM algorithm is only about $5\%$. This result shows that our proposed learning-based solutions yield a good balance between system performance and feasibility for practical implementation.

\begin{figure}
\setlength{\abovecaptionskip}{0.3cm plus 0pt minus 0pt}
\begin{subfigure}{.5\textwidth}
\centering
\includegraphics[scale=0.23]{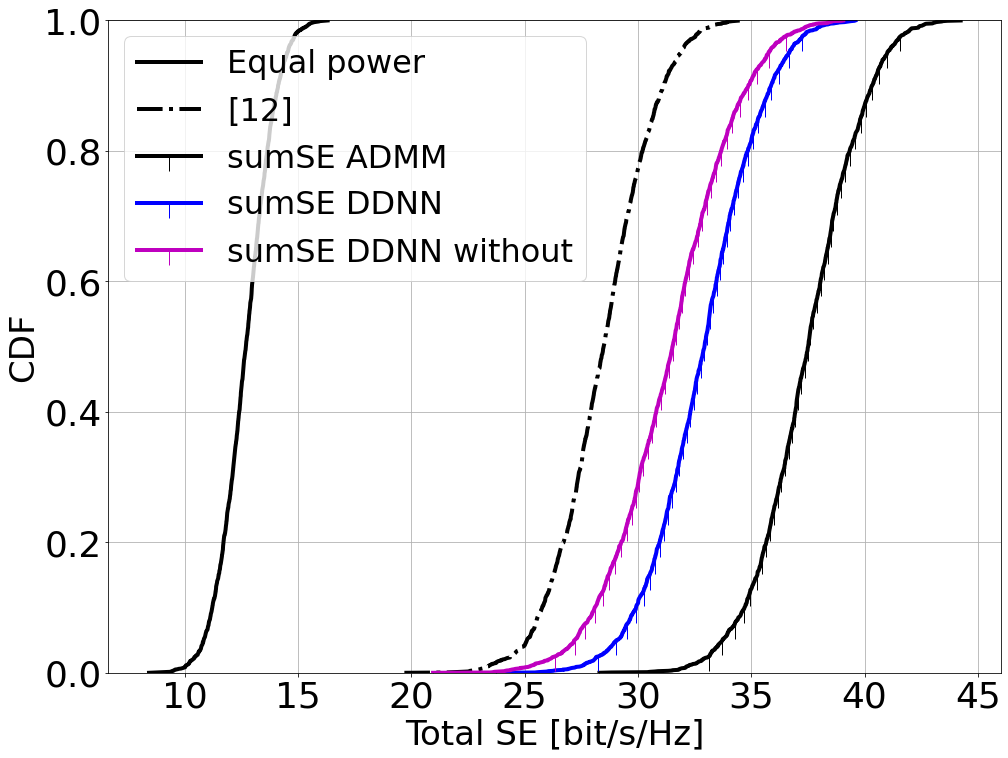}
\caption{MR.}
\label{totalSE_MR}
\vspace{1em}
\end{subfigure}
\begin{subfigure}{.5\textwidth}
\centering
\includegraphics[scale=0.23]{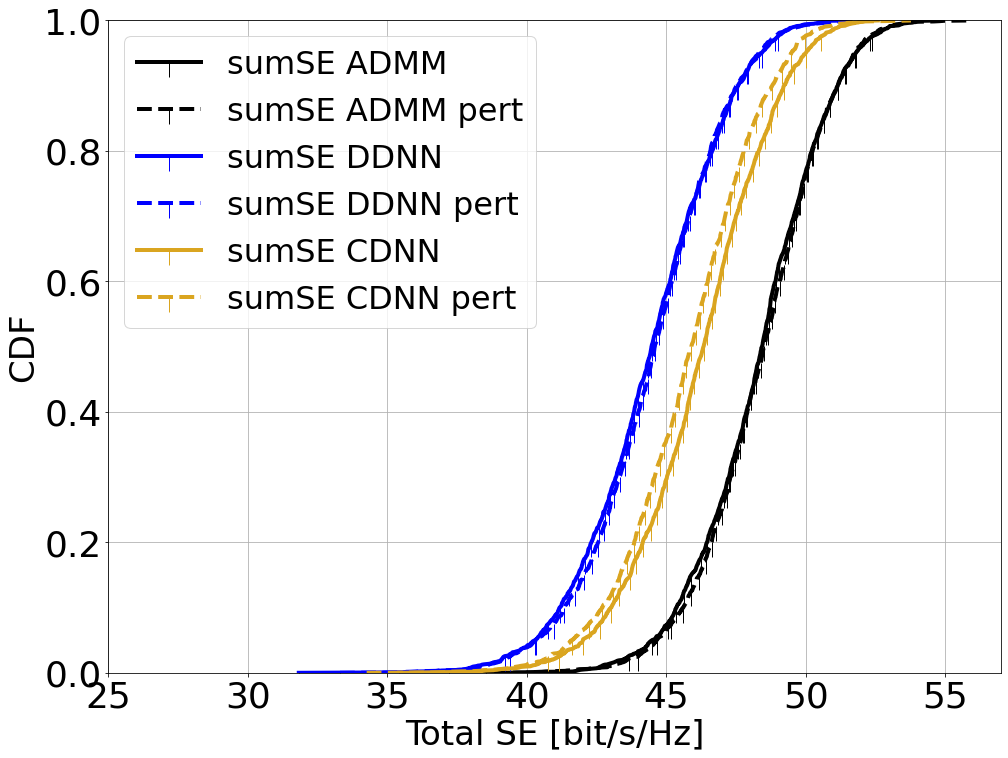}
\caption{RZF.}
\label{totalSE_pert}
\end{subfigure}
\caption{CDF of the total DL SE for sum-SE maximization objective.}
\label{totalSE_robustness}
\vspace{-0.5em}
\end{figure}

Fig. \ref{totalSE_MR} plots the CDF of the total DL SE for sum-SE maximization with MR precoding for the DDNN model with and without the heuristic pre-processing and total power approximation. It can be seen that the proposed pre/post-processing provides up to $8 \%$ enhancement in the total DL SE, at a negligible cost in terms of computational complexity, increasing the performance improvement over \cite{interdonato2019scalability}. Further, to demonstrate the robustness of the learning-based solutions, Fig. \ref{totalSE_pert} compares the CDF of the total DL SE for the original DDNN and CDNN models with perturbed LSF inputs. A truncated log-normal perturbation with a standard deviation of $1$\,dB, and truncated at $3$\,dB is chosen to model small variations in the propagation environment (for example due to vehicles and trees). Note that the DNNs are not retrained with the perturbed input in order to test their robustness against unknown variations that may occur in a practical cell-free network deployment. It is clear that the input perturbation does not have a noticeable effect on the fully distributed DDNN model, whereas the clustered model shows a minor degradation in performance. The reason behind this is that the clustered model has a larger input vector including the LSF coefficients of the UEs to different APs, and so the perturbations here affect as well the inter-relations between the LSF coefficients of a given UE to the APs in the cluster.

\section{Conclusions}\label{conc}

In this paper, the power allocation problem in the DL of a cell-free massive MIMO system has been addressed. MR and RZF precoding have been considered. First, we formulate the network-wide sum-SE and PF maximization problems and utilize the WMMSE-ADMM algorithm to generate a large amount of data in a reasonable time. We then propose different distributed and clustered DNNs to approximate the power coefficients using only the LSF coefficients that are locally available at an AP/EP, and so allowing for reduced fronthaul/midhaul requirements. We draw attention to that our DNN models find an unknown mapping from a smaller domain input, than the conventional WMMSE-ADMM algorithm, to the desired output power coefficients. Moreover, the run-time for the learning-based algorithms has been shown to be much less than that of the ADMM algorithm, which is the most efficient conventional approach for solving the power allocation optimization problem, to the best of our knowledge. The reduced fronthaul and computational time indicate that our proposed models are particularly useful for implementation in large-scale systems. Numerical results show that our distributed and clustered DNN implementations outperform low-complexity practical heuristic allocations and provide a reasonable approximation to the (centralized) WMMSE-ADMM benchmark.

\section*{References}
\renewcommand{\refname}{ \vspace{-\baselineskip}\vspace{-1.1mm} }
\bibliographystyle{ieeetr}
\bibliography{papercites}

\begin{thebibliography}{10}

\bibitem{marzetta2016fundamentals}
T.~L. Marzetta, E.~G. Larsson, H.~Yang, and H.~Q. Ngo, {\em Fundamentals of
  massive MIMO}.
\newblock Cambridge University Press, 2016.

\bibitem{bjornson2017book}
E.~Bj{\"o}rnson, J.~Hoydis, and L.~Sanguinetti, ``Massive {MIMO} networks:
  Spectral, energy, and hardware efficiency,'' {\em Foundations and Trends in
  Signal Processing}, vol.~11, no.~3-4, pp.~154--655, 2017.

\bibitem{bjornson2017massive}
E.~Bj{\"o}rnson, J.~Hoydis, and L.~Sanguinetti, ``Massive {MIMO} has unlimited
  capacity,'' {\em IEEE Transactions on Wireless Communications}, vol.~17,
  no.~1, pp.~574--590, 2017.

\bibitem{nguyen2018optimal}
T.~H. Nguyen, T.~K. Nguyen, H.~D. Han, {\em et~al.}, ``Optimal power control
  and load balancing for uplink cell-free multi-user massive {MIMO},'' {\em
  IEEE access}, vol.~6, pp.~14462--14473, 2018.

\bibitem{Ngo2017b}
H.~Q. Ngo, A.~Ashikhmin, H.~Yang, E.~G. Larsson, and T.~L. Marzetta,
  ``Cell-free {Massive} {MIMO} versus small cells,'' {\em IEEE Trans. Wireless
  Comm.}, vol.~16, no.~3, pp.~1834--1850, 2017.

\bibitem{zhang2019cell}
J.~Zhang, S.~Chen, Y.~Lin, J.~Zheng, B.~Ai, and L.~Hanzo, ``Cell-free massive
  {MIMO}: A new next-generation paradigm,'' {\em IEEE Access}, vol.~7,
  pp.~99878--99888, 2019.

\bibitem{buzzi2017downlink}
S.~Buzzi and A.~Zappone, ``Downlink power control in user-centric and cell-free
  massive {MIMO} wireless networks,'' in {\em 2017 IEEE 28th Annual
  International Symposium on Personal, Indoor, and Mobile Radio Communications
  (PIMRC)}, pp.~1--6, IEEE, 2017.

\bibitem{bjornson2019making}
E.~Bj{\"o}rnson and L.~Sanguinetti, ``Making cell-free massive {MIMO}
  competitive with {MMSE} processing and centralized implementation,'' {\em
  IEEE Transactions on Wireless Communications}, vol.~19, no.~1, pp.~77--90,
  2019.

\bibitem{chakraborty2020efficient}
S.~Chakraborty, {\"O}.~T. Demir, E.~Bj{\"o}rnson, and P.~Giselsson, ``Efficient
  downlink power allocation algorithms for cell-free massive {MIMO} systems,''
  {\em IEEE Open Journal of the Communications Society}, vol.~2, pp.~168--186,
  2021.

\bibitem{demir2021foundations}
{\"O}.~T. Demir, E.~Bj{\"o}rnson, and L.~Sanguinetti, ``Foundations of
  user-centric cell-free massive {MIMO},'' {\em Foundations and
  Trends{\textregistered} in Signal Processing}, vol.~14, no.~3-4,
  pp.~162--472, 2021.

\bibitem{Ngo2018a}
H.~Q. {Ngo}, L.~{Tran}, T.~Q. {Duong}, M.~{Matthaiou}, and E.~G. {Larsson},
  ``On the total energy efficiency of cell-free massive {MIMO},'' {\em IEEE
  Trans. Green Commun. Net.}, vol.~2, no.~1, pp.~25--39, 2018.

\bibitem{interdonato2019scalability}
G.~Interdonato, P.~Frenger, and E.~G. Larsson, ``Scalability aspects of
  cell-free massive {MIMO},'' in {\em IEEE International Conference on
  Communications (ICC)}, pp.~1--6, IEEE, 2019.

\bibitem{zhang2020prospective}
J.~Zhang, E.~Bj{\"o}rnson, M.~Matthaiou, D.~W.~K. Ng, H.~Yang, and D.~J. Love,
  ``Prospective multiple antenna technologies for beyond {5G},'' {\em IEEE
  Journal on Selected Areas in Communications}, vol.~38, no.~8, pp.~1637--1660,
  2020.

\bibitem{interdonato2019ubiquitous}
G.~Interdonato, E.~Bj{\"o}rnson, H.~Q. Ngo, P.~Frenger, and E.~G. Larsson,
  ``Ubiquitous cell-free massive {MIMO} communications,'' {\em EURASIP Journal
  on Wireless Communications and Networking}, vol.~2019, no.~1, pp.~1--13,
  2019.

\bibitem{burr2018cooperative}
A.~Burr, M.~Bashar, and D.~Maryopi, ``Cooperative access networks: Optimum
  fronthaul quantization in distributed massive {MIMO} and cloud {RAN}-invited
  paper,'' in {\em IEEE 87th Vehicular Technology Conference (VTC Spring)},
  pp.~1--5, 2018.

\bibitem{buzzi2019user}
S.~Buzzi, C.~D’Andrea, A.~Zappone, and C.~D’Elia, ``User-centric {5G}
  cellular networks: Resource allocation and comparison with the cell-free
  massive {MIMO} approach,'' {\em IEEE Transactions on Wireless
  Communications}, vol.~19, no.~2, pp.~1250--1264, 2019.

\bibitem{masoumi2019performance}
H.~Masoumi and M.~J. Emadi, ``Performance analysis of cell-free massive {MIMO}
  system with limited fronthaul capacity and hardware impairments,'' {\em IEEE
  Transactions on Wireless Communications}, vol.~19, no.~2, pp.~1038--1053,
  2019.

\bibitem{bashar2020exploiting}
M.~Bashar, A.~Akbari, K.~Cumanan, H.~Q. Ngo, A.~G. Burr, P.~Xiao, M.~Debbah,
  and J.~Kittler, ``Exploiting deep learning in limited-fronthaul cell-free
  massive {MIMO} uplink,'' {\em IEEE Journal on Selected Areas in
  Communications}, vol.~38, no.~8, pp.~1678--1697, 2020.

\bibitem{femenias2020fronthaul}
G.~Femenias and F.~Riera-Palou, ``Fronthaul-constrained cell-free massive
  {MIMO} with low resolution {ADCs},'' {\em IEEE Access}, vol.~8,
  pp.~116195--116215, 2020.

\bibitem{sarajlic2019fully}
M.~Sarajli{\'c}, F.~Rusek, J.~R. S{\'a}nchez, L.~Liu, and O.~Edfors, ``Fully
  decentralized approximate zero-forcing precoding for massive {MIMO}
  systems,'' {\em IEEE Wireless Communications Letters}, vol.~8, no.~3,
  pp.~773--776, 2019.

\bibitem{jeon2019decentralized}
C.~Jeon, K.~Li, J.~R. Cavallaro, and C.~Studer, ``Decentralized equalization
  with feedforward architectures for massive {MU-MIMO},'' {\em IEEE
  Transactions on Signal Processing}, vol.~67, no.~17, pp.~4418--4432, 2019.

\bibitem{bjornson2020scalable}
E.~Bj{\"o}rnson and L.~Sanguinetti, ``Scalable cell-free massive {MIMO}
  systems,'' {\em IEEE Transactions on Communications}, vol.~68, no.~7,
  pp.~4247--4261, 2020.

\bibitem{zhao2020power}
Y.~Zhao, I.~G. Niemegeers, and S.~H. De~Groot, ``Power allocation in cell-free
  massive {MIMO}: A deep learning method,'' {\em IEEE Access}, vol.~8,
  pp.~87185--87200, 2020.

\bibitem{nayebi2017precoding}
E.~Nayebi, A.~Ashikhmin, T.~L. Marzetta, H.~Yang, and B.~D. Rao, ``Precoding
  and power optimization in cell-free massive {MIMO} systems,'' {\em IEEE
  Transactions on Wireless Communications}, vol.~16, no.~7, pp.~4445--4459,
  2017.

\bibitem{goodfellow2016deep}
I.~Goodfellow, Y.~Bengio, A.~Courville, and Y.~Bengio, {\em Deep learning},
  vol.~1.
\newblock MIT press Cambridge, 2016.

\bibitem{lee2018deep}
W.~Lee, M.~Kim, and D.-H. Cho, ``Deep power control: Transmit power control
  scheme based on convolutional neural network,'' {\em IEEE Communications
  Letters}, vol.~22, no.~6, pp.~1276--1279, 2018.

\bibitem{sun2018learning}
H.~Sun, X.~Chen, Q.~Shi, M.~Hong, X.~Fu, and N.~D. Sidiropoulos, ``Learning to
  optimize: Training deep neural networks for interference management,'' {\em
  IEEE Transactions on Signal Processing}, vol.~66, no.~20, pp.~5438--5453,
  2018.

\bibitem{sanguinetti2018deep}
L.~Sanguinetti, A.~Zappone, and M.~Debbah, ``Deep learning power allocation in
  massive {MIMO},'' in {\em 52nd Asilomar conference on signals, systems, and
  computers}, pp.~1257--1261, 2018.

\bibitem{chakraborty2019centralized}
S.~Chakraborty, E.~Bj{\"o}rnson, and L.~Sanguinetti, ``Centralized and
  distributed power allocation for max-min fairness in cell-free massive
  {MIMO},'' in {\em 53rd Asilomar Conference on Signals, Systems, and
  Computers}, pp.~576--580, 2019.

\bibitem{zhang2019deep}
C.~Zhang, P.~Patras, and H.~Haddadi, ``Deep learning in mobile and wireless
  networking: A survey,'' {\em IEEE Communications Surveys \& Tutorials},
  vol.~21, no.~3, pp.~2224--2287, 2019.

\bibitem{van2020power}
T.~Van~Chien, T.~N. Canh, E.~Bj{\"o}rnson, and E.~G. Larsson, ``Power control
  in cellular massive {MIMO} with varying user activity: A deep learning
  solution,'' {\em IEEE Transactions on Wireless Communications}, vol.~19,
  no.~9, pp.~5732--5748, 2020.

\bibitem{nasir2019multi}
Y.~S. Nasir and D.~Guo, ``Multi-agent deep reinforcement learning for dynamic
  power allocation in wireless networks,'' {\em IEEE Journal on Selected Areas
  in Communications}, vol.~37, no.~10, pp.~2239--2250, 2019.

\bibitem{zhang2020multi}
L.~Zhang and Y.-C. Liang, ``Multi-agent deep reinforcement learning for
  non-cooperative power control in heterogeneous networks,'' in {\em GLOBECOM
  2020-2020 IEEE Global Communications Conference}, pp.~1--6, IEEE, 2020.

\bibitem{de2018team}
P.~de~Kerret, D.~Gesbert, and M.~Filippone, ``Team deep neural networks for
  interference channels,'' in {\em 2018 IEEE International Conference on
  Communications Workshops (ICC Workshops)}, pp.~1--6, IEEE, 2018.

\bibitem{christensen2008weighted}
S.~S. Christensen, R.~Agarwal, E.~De~Carvalho, and J.~M. Cioffi, ``Weighted
  sum-rate maximization using weighted {MMSE} for {MIMO-BC} beamforming
  design,'' {\em IEEE Transactions on Wireless Communications}, vol.~7, no.~12,
  pp.~4792--4799, 2008.

\bibitem{shi2011iteratively}
Q.~Shi, M.~Razaviyayn, Z.-Q. Luo, and C.~He, ``An iteratively weighted {MMSE}
  approach to distributed sum-utility maximization for a {MIMO} interfering
  broadcast channel,'' {\em IEEE Transactions on Signal Processing}, vol.~59,
  no.~9, pp.~4331--4340, 2011.

\bibitem{diamantoulakis2017maximizing}
P.~D. Diamantoulakis and G.~K. Karagiannidis, ``Maximizing proportional
  fairness in wireless powered communications,'' {\em IEEE Wireless
  Communications Letters}, vol.~6, no.~2, pp.~202--205, 2017.

\bibitem{chen2019proportional}
L.~Chen, L.~Ma, and Y.~Xu, ``Proportional fairness-based user pairing and power
  allocation algorithm for non-orthogonal multiple access system,'' {\em IEEE
  Access}, vol.~7, pp.~19602--19615, 2019.

\bibitem{demir2020lsfp}
{\"O}.~T. Demir and E.~Bj{\"o}rnson, ``Large-scale fading precoding for
  spatially correlated {R}ician fading with phase shifts,'' {\em 2020,
  arXiv:2006.14267. [Online]. Available: https://arxiv.org/abs/2006.14267}.

\bibitem{browncloud}
G.~Brown, ``Cloud-{RAN}, the next-generation mobile network architecture,''
  {\em 2017, Huawei White Paper. [Online]. Available:
  https://www-file.huawei.com/-/media/CORPORATE/PDF/mbb/cloud-ran-the-next-generation-mobile-network-architecture.pdf?la=en}.

\bibitem{sriram2019joint}
A.~Sriram, M.~Masoudi, A.~Alabbasi, and C.~Cavdar, ``Joint functional splitting
  and content placement for green hybrid {CRAN},'' in {\em IEEE 30th Annual
  International Symposium on Personal, Indoor and Mobile Radio Communications
  (PIMRC)}, pp.~1--7, 2019.

\bibitem{habibi2019comprehensive}
M.~A. Habibi, M.~Nasimi, B.~Han, and H.~D. Schotten, ``A comprehensive survey
  of {RAN} architectures toward {5G} mobile communication system,'' {\em IEEE
  Access}, vol.~7, pp.~70371--70421, 2019.

\end{thebibliography}

\end{document}